\documentclass[a4paper,fleqn,usenatbib]{mnras}
\usepackage{soul}
\usepackage[T1]{fontenc}
\usepackage{ae,aecompl}
\usepackage{deluxetable}
\usepackage{subfig}
\usepackage[dvipdfm]{graphicx}	
\usepackage{mwe}
\usepackage{amsmath}	
\let\oldAA\AA
\renewcommand{\AA}{\text{\normalfont\oldAA}}
\newcommand{\teff}{T_\mathrm{eff}}
\newcommand{\logg}{\mathrm{log}\,g}
\usepackage{graphicx}
\usepackage{natbib}
\usepackage{ulem}
\usepackage{subfig}
\usepackage{float}
\usepackage[usenames,dvipsnames]{xcolor}
\usepackage{multirow}
\usepackage{footnote}
\usepackage[bottom]{footmisc}
\usepackage{xltabular}
\usepackage{longtable}
\usepackage{lipsum}
\usepackage{pdflscape}
\usepackage{pdfpages}
\usepackage[export]{adjustbox}

\newcommand{\ha}{H$\alpha$\ }
\newcommand{\hb}{H$\beta$\ }

\title[H$\alpha$ variability in M dwarfs]{Exploring the short-term variability of H$\alpha$ and H$\beta$ emissions in a sample of M dwarfs}

\author[Kumar et al.]{Vipin Kumar$^{1,2,3}$\thanks{kumar@ph1.uni-koeln.de},
	A. S. Rajpurohit$^{1}$\thanks{arvindr@prl.res.in},
	Mudit K. Srivastava$^{1}$\thanks{mudit@prl.res.in},
	\newauthor Jos\'e G. Fern\'andez-Trincado$^{4}$ and
	A. B. A. Queiroz$^{5,6,7}$
	\\
	$^{1}$Astronomy \& Astrophysics Division, Physical Research Laboratory, Ahmedabad 380009, India\\
	$^{2}$Department of Physics, Indian Institute of Technology, Gandhinagar, 382335, India\\
	$^{3}$I. Physikalisches Institut, Universit\"at zu K\"oln, Z\"ulpicher Stra\ss{}e 77, 50937, K\"oln, Germany\\
	$^{4}$Instituto de Astronom\'ia, Universidad Cat\'olica del Norte, Av. Angamos 0610, Antofagasta, Chile\\
	$^{5}$Leibniz-Institut f\"ur Astrophysik Potsdam (AIP), An der Sternwarte 16, 14482 Potsdam, Germany\\
	$^{6}$Institut f\"{u}r Physik und Astronomie, Universit\"{a}t Potsdam, Haus 28 Karl-Liebknecht-Str. 24/25, D-14476 Golm, Germany\\
	$^{7}$Laborat\'orio Interinstitucional de e-Astronomia - LIneA, Rua Gal. Jos\'e Cristino 77, Rio de Janeiro, RJ - 20921-400, Brazil
}

\begin{document}
	
\date{Accepted YYYY Month DD.  Received YYYY Month DD; in original form YYYY Month DD}
	
\maketitle
	
\label{firstpage}	

\begin{abstract}	
	
The time scales of variability in active M dwarfs can be related to their various physical parameters. Thus, it is important to understand such variability to decipher the physics of these objects. In this study, we have performed the low resolution ($\sim$5.7 \AA) spectroscopic monitoring of 83 M dwarfs (M0-M6.5) to study the variability of H$\alpha$ / H$\beta$ emissions; over the time scales from $\sim$0.7 to 2.3 hours with a cadence of $\sim$3-10 minutes. Data of a sample of another 43 late-type M dwarfs (M3.5-M8.5) from the literature are also included to explore the entire spectral sequence. 53 of the objects in our sample ($\sim$64\%) show statistically significant short-term variability in H$\alpha$. We show that this variability in 38 of them are most likely to be related to the flaring events. We find that the early M dwarfs are less variable despite showing higher activity strengths (L$_{H\alpha}$/L$_{bol}$ \& L$_{H\beta}$/L$_{bol}$), which saturates around $\sim$10$^{-3.8}$ for M0-M4 types. Using archival photometric light curves from TESS and Kepler/K2 missions, the derived chromospheric emission (\ha and \hb emission) variability is then explored for any plausible systematics with respect to their rotation phase. The variability indicators clearly show higher variability in late-type M dwarfs (M5-M8.5) with shorter rotation periods ($<$2 days). For 44 sources, their age has been estimated using StarHorse project and possible correlations with variability have been explored. The possible causes and implications for these behaviors are discussed.

\end{abstract}


\section{Introduction}
\label{sec-intro}

M dwarfs are the major stellar constituents of the Galaxy. Their population is estimated to be nearly $\sim$70 \% of the total stellar content in our Galaxy, and they contribute $\sim$40\% of its total stellar mass \citep{Henry1997, Chabrier2003}. M dwarfs are less massive (0.6-0.075 M$_{\odot}$) and cooler stars with effective temperature ($\teff$) in the range of 2500-4000 K. As their mean lifetimes on the main sequence path are comparatively much longer, they are a good tracer of the Galactic history \citep{Green1994, Cool1996, Renzini1996}. A variety of them are expected to host sub-stellar objects, e.g., brown dwarfs and exoplanets, \citep{Bonfils2012, Gillon2017, Mercer2020, Baroch2021}, therefore they have been a major topic of attention in recent times. 

\par
It is well established that magnetic fields are the sole reason for stellar magnetic activity and almost certainly play a fundamental role in the physics of late-type stellar atmospheres. A large fraction of M dwarfs are known to be magnetically active \citep{West2008, West2015}.  It is expected that 30-40 $\%$ of M dwarfs are members of stellar or sub-stellar binary systems. Interactions between M dwarfs and their binary companion may impact the companion’s evolution due to the activity of the M dwarf as suggested by \cite{Kouwenhoven2009} and \cite{Dressing2015}.

\par

M dwarfs with masses below 0.35M$_{\odot}$ \citep{Chabrier1997}, become fully convective and, unlike solar-type stars, lack the tachocline region, which is thought to be essential for their magnetic field generation. In such fully convective M dwarfs, the mechanisms for generating large-scale magnetic fields are not yet well understood \citep{Newton2017}. Therefore, a thorough understanding of the M dwarf magnetic activity is essential to explore various physical processes associated with its generation.

\par

Observable phenomena produced in the outer stellar atmosphere, such as strong stellar winds, flares, coronal mass ejection, spots, etc., are used to describe the level of magnetic activity in stars. The magnetic heating of the stellar atmosphere results in various chromospheric emissions such as \textit{Ca} \textit{II} H+K, \textit{Na} \textit{I} D, \textit{Mg} \textit{II} and \textit{K} along with H$\alpha$ are commonly used as a proxy of magnetic activity in M dwarfs \citep{Hawley1996, Lee2010, Fuhrmeister2019, Schoefer2019}. Out of these, the chromospheric H$\alpha$ emission line is widely used for activity-related studies, as it is easily observable in M dwarfs compared to other lines in the faint blue part of the spectrum \citep{Walkowicz2009A}. Because of magnetic activity, these atomic lines can be affected by line profile changes which are thought to be caused by star spots or plages \citep{Schoefer2019}.

\par

Magnetic activity indicators, such as H$\alpha$ emission in the chromosphere, and X-ray emissions in the corona, is measurable evidence of their surface magnetism. Studies by \cite{Angus2015, West2008, Mamajek2008, Reiners2012, Newton2017, Riedel2017, Kiman2019} and \cite{Kiman2021} show that magnetic activity is known to be correlated with stellar age. These magnetic activity are partly responsible for the stellar magnetic wind, which dissipates angular momentum, and appears to scale with stellar rotation. Thus magnetic activity and age are tightly related for solar-type stars \citep{Mamajek2008}. Furthermore, \cite{Reiners2012}, \cite{West2015}, \cite{Newton2017}, and \cite{Riedel2017} find that in M dwarfs, the magnetic activity decreases with age for late-M dwarfs though, for fully convective M dwarfs, the mechanism to generate magnetic fields is not yet well understood.

\par

The tracers of magnetic activity are also closely tied to stellar rotation in solar-type stars and become stronger for stars, which rotate faster \citep{Pallavicini1981, Wright2011, Reiners2014}. The stellar rotation period can be determined with various spectroscopic and photometric measurements. \cite{Suarez2018}, and  \cite{Fuhrmeister2019} measured the rotation periods of early M dwarfs using spectroscopic indicators such as H$\alpha$, \textit{Ca} \textit{II} H+K, or \textit{K} where the inhomogeneous distributions of stellar surface features lead to variations over the course of its rotation. While \cite{Kiraga2007, Irwin2011, Newton2016, Newton2018} measured the rotation periods for M dwarfs using photometry by measuring the brightness variations caused by long-lived star spots. 

\par

A strong correlation between rotation period and magnetic activity is found for early-type M dwarfs (masses $>$0.35 M$_{\odot}$) \citep{Reiners2012, West2015, Newton2017}. M dwarfs earlier than M3 show a clear saturated rotation-activity relationship as demonstrated in coronal X-ray emission \citep{Wright2011}, and H$\alpha$, \textit{Ca} \textit{II} H+K emission in the chromosphere. Recently, \cite{Wright2018} finds that fully convective M dwarfs  (masses $<$0.35 M$_{\odot}$) also follow the same X-ray rotation–activity relationship. Further, \cite{Newton2017} suggested that depending on the mass, M dwarfs with spectral type earlier than M2.5 shows an increment in H$\alpha$ activity with a decreasing period and reaches a saturated level for Rossby numbers (a quantity which describes the strength of the rotational effect on the convective flows) smaller than $\sim$ 0.1 and start to decline after R$_0$ $>$ 0.2.

\par

\cite{West2015} shows that in M dwarfs, the activity fraction decreases with increasing rotation period for early-type M dwarfs (M1-M4), whereas the higher activity fraction for late-type M dwarfs extends to slower rotation rates. {Similarly, \cite{Jenkins2009} finds the change in the rotation period at the fully convective boundary. They attribute this to the changing field topology between partially and fully convective stars as suggested by \citep{Reiners2007}. However, with similar rotation periods and masses in fully convective M dwarfs, \cite{Morin2010} found that magnetic field topologies fall largely into two categories: one group has strong, axisymmetric, largely dipolar global fields, while the other group showed weak, non-axisymmetric global fields.

\par

In recent times, TESS and Kepler/K2 missions \citep{Caldwell2010, Koch2010, Howell2014, Ricker2015} have offered another opportunity to study various activity indicators, e.g., star-spots, bright faculae, etc. The high cadence data from these missions are successfully used to explore the short-duration activity in the photometry light curves \citep{Doyle2018, Doyle2019}. The modulation in the photometric light curve originates due to the presence of starspots/bright faculae on the stellar surface \citep{Radick1998, Hall2009, Buccino2011}. Thus the correlations/anti-correlation of any activity indicator strength (such as \ha emission) with the rotation phase have also been examined in the context of emission originating from such star spots and/or bright faculae. Such previous studies e.g., \cite{Radick1998, Hall2009, Buccino2011, Medina2022} indicate that some M dwarfs show a correlation/anti-correlation between chromospheric emission strength with the rotation phase. i.e., the emission strength increases with the increase/decrease in brightness of the star (caused by the rotation of the star).

\par 
The magnetic activity in the M dwarf happens at various time scales ranging from a few seconds to several hours \citep{Kowalski2010, Yang2017}. \cite{Doyle2018, Doyle2019} demonstrated that magnetic activity and energetic flaring events on the stellar surface could vanish in seconds to hours. Such activity would then invariably be seen in the flux variation in H$\alpha$ \citep{Lee2010, Hilton2010, Almeida2011, Walkowicz2011, Hawley2014, Chang2017}. While a good sample of M dwarfs has been studied to characterise the activity at the larger time scales through H$\alpha$ variability, very few studies in the literature investigate H$\alpha$ variability at shorter time scales of a few minutes (for e.g., 5-20 minutes and/or with a sample of uneven cadence). Thus, short-duration behaviour could not be probed, leading to a gap in the systematic understanding of  H$\alpha$ variability on such time scales.

\par

One such short-scale systematic data set was presented by \cite{Lee2010}, who had studied 43 sources of M3.5-M8.5 spectral range at a cadence of $\sim$5 to 10 minutes over a timescale of $\sim$ 0.1-1 hr. Though later \cite{Kruse2010} had expanded this study over a complete spectral sequence (M0-M9), they mostly used SDSS survey data of longer exposure time ($\sim$15 minutes). Recently, \cite{Medina2022} used high-cadence spectroscopic data to examine H$\alpha$ variability on the timescales of minutes to hours on a sample of 13 fully convective, active mid-to-late M stars. Their study concludes that the dominant source of H$\alpha$ variability on the timescales could be the low-energy flares.

\par

Therefore, in this study, we have performed a systematic short-term (mostly 5 minutes individual frame exposures over 0.7-2.3 hours) spectroscopy monitoring of a sample of M dwarfs in the spectral range of M0-M6.5 to probe their \ha variability. This chosen spectral range was apt as our sample of 83 M dwarfs in M0-M6.5 spectral class complemented the data set provided by \cite{Lee2010} in the range of M3.5-M8.5. We have, thus, constructed a sample of $\sim126$ sources in the complete spectral range of M0-M8.5.

\par

The outline of this paper is as follows. In Section \ref{sec-ObsMtAbu}, we have described the sample selection criteria and observations. In Section \ref{sec-spec_result}, we have discussed the results from our spectroscopic analysis, like the estimation of atmospheric parameters, \ha and \hb variability indicators, activity strength measurements, etc. In Section \ref{sec-phot_result}, we have estimated the rotation periods and star-spot filling factor with the help of photometric light curves and relate these rotation periods with the variability indicators. In Section \ref{sec-age}, we have derived the age of the sources using StarHorse project and discussed observed variability with their age. Finally, we summarized our results and discussed the implications in Section~\ref{Sec-Discussion}.

\section{Sample selection \& Observations}
\label{sec-ObsMtAbu}
For this observing campaign, we targeted the M dwarfs in the spectral range M0-M6.5 with MFOSC-P instrument on the PRL 1.2 m, f/13 telescope \citep{Srivastava2018,Srivastava2021, Rajpurohit2020}. The moderate aperture of the telescope and cumulative (telescope + instrument) efficiency compelled us to restrict our sample with V magnitude brighter than 14 typically. This also restricts us from expanding our spectral range beyond M6 as most of the late M dwarf (M7 and beyond) are fainter (V$>$16) for spectroscopy with MFOSC-P on 1.2m telescope. The distribution of all the sources is shown in Fig.~\ref{figure1}. The selected sources in this study have typical \ha equivalent widths (EW) $<$ -0.75 $\AA$, which usually corresponds to the detectable \ha line in emission. Thus, we selected 83 suitable targets from the list of \cite{Jeffers2018} and \cite{lepine2011}. They were observed between March 2020 to  March 2021. The details of these targets are summarized in Table~\ref{table-ObsSpecMFOSCP1}.
\par
The instrument MFOSC-P (Mount-Abu Faint Object Spectrograph and Camera-Pathfinder) is an imager-spectrograph \citep{Srivastava2018,Srivastava2021} which provides visible-band spectroscopy using three plane reflection gratings of 600, 300, and 150 line-pairs(lp) mm$^{-1}$. These gratings offer resolutions of R$\sim$2000, 1000 and 500 centered at $\sim$6500, 5500, 6000\AA~ and are referred to as R2000, R1000, and R500 modes, respectively. For the current observing program, we have utilized R1000 mode (300 lp mm$^{-1}$, dispersion $\sim$1.9$\AA$ per pixel) with 1 arc-second ($\sim$3 pixels) slit-width covering a spectral range of 4700-6650\AA~, i.e., covering both \hb and \ha wavelengths. The targets were monitored with integration times in the range of 200-600s per frame for $\sim$0.7-2.3 hours in a single stretch for each of the sources. Thus, each data set consists of typically $\sim$8-18 frames of the individual spectrum. A lower-resolution spectrum was also recorded for each of the sources in R500 mode (150 lp mm$^{-1}$, dispersion $\sim$3.8$\AA$ per pixel) to cover a larger wavelength range $\sim$4500-8500\AA~ for spectral classification purposes. Xenon lamp spectra for wavelength calibration were obtained at the beginning/end of each spectral time series in the identical settings of the instrument. Spectro-photometric standard stars from the ESO catalog \footnote{https://www.eso.org/sci/observing/tools/standards/spectra/stanlis.html} were observed (on the same or contemporaneous nights) in the identical setting of the instrument to correct the instrument response. Subsequently, the standard MFOSC-P data reduction procedure has been applied to produce the science-ready spectra (see \cite{Rajpurohit2020} for details). The log of the MFOSC-P spectroscopy are given in Table~\ref{table-ObsSpecMFOSCP1}.Photometric light curves of 75 of the above sources were obtained from the TESS and Kepler/K2 archival databases through Mikulski Archive for Space Telescopes (MAST) portal\footnote{https://mast.stsci.edu/portal/Mashup/Clients/Mast/
	
	Portal.html}. The related analysis and derived results are discussed in Section~\ref{sec-phot_result}.
\par

\begin{figure}
	\centering
	\includegraphics[angle=0,width=0.50\textwidth]{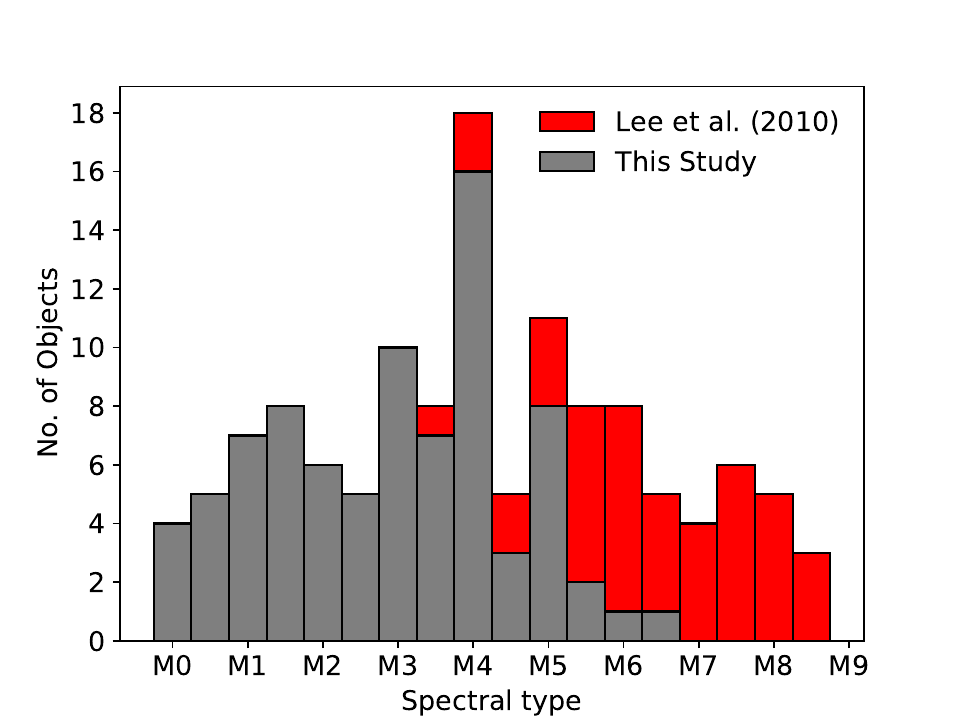}
	\caption{Distribution of 83 M dwarfs of this study along with 43 M dwarfs from \protect \cite{Lee2010} with respect to the spectral type.}
	\label{figure1}
\end{figure}

\section{Analysis and Results from Spectroscopy}
\label{sec-spec_result}

\subsection{Determination of Atmospheric Parameters and Spectral Class}

Though the spectral classification of the targets is given in  \cite{lepine2013} and \cite{Jeffers2018}, we nevertheless choose to re-confirm those with MFOSC-P low resolution (R500 mode) spectra. Even though the second-order contamination (beyond 7600\AA~) is expected to be minimal given the low U- to I-band flux ratios for M dwarfs (as well as the low spectral throughput of the instrument $+$ telescope in the bluer part), we have restricted the spectral range to 4500-7500 $\AA$ for this purpose. The spectral classification was derived by comparing the spectra with M dwarf templates from \cite{Bochanski2007} using a similar approach as adopted in \cite{Rajpurohit2020}. The derived spectral classes are in good agreement with the spectral classes given by \cite{lepine2013}, and \cite{Jeffers2018} within one spectral class. In this work, however, we shall be using the spectral classes of \cite{lepine2013}, and \cite{Jeffers2018} as they were derived with higher resolution spectra. The spectral classes for six sources were not available in these references. Thus, we took their values from our analysis.
\par
To derive the atmospheric parameters $\teff$, surface gravity ($\logg$), the observed spectra (R$\sim$500) were compared with the BT-Sett synthetic spectra  \citep{Allard2010,Allard2013, Rajpurohit2012a}, similar to the approach adopted in \cite{Rajpurohit2020}. The grid for $\teff$ spans between 3000 to 4000 K in steps of 100 K, and $\logg$ ranges from 4.0 to 5.5 in steps of 0.5 dex has been used for the comparison. Similar to \cite{Rajpurohit2020}, the grid at solar metallicity is used as the sources in this study also lies within 100 pc of the solar neighborhood. In Fig.~\ref{figure2}, we show one such comparison of the observed spectra with the synthetic ones for the spectral range M0-M6.5. The spectral range between 5000 to 7500 $\AA$ has been used for the $\chi^2$ minimization. The spectral regions between 6540-6585 $\AA$ and 6840-6920 $\AA$ (containing \ha and \hb emission line and telluric feature, respectively) were excluded for this analysis. The parameters derived would, thus, have errors equal to the grid spacing, i.e., 100 K for $\teff$ and 0.5 dex for $\logg$. The derived atmospheric parameters of all the sources are summarized in Table~\ref{table-ObsSpecMFOSCP1} along with the observing log. 


\begin{figure}
	\centering
	\includegraphics[angle=0,width=0.48\textwidth]{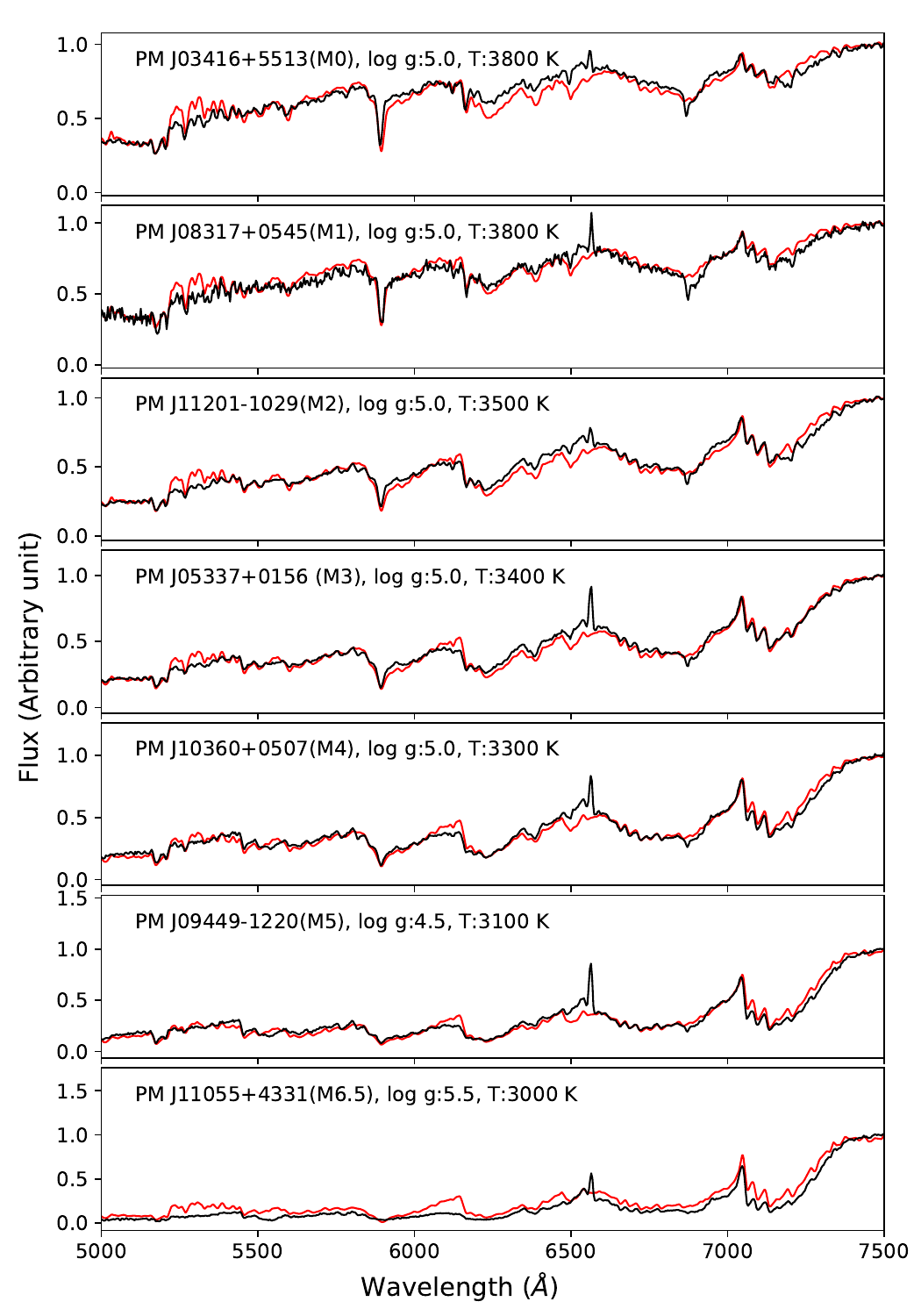}
	\caption{Comparison of observed spectra of M dwarfs (black) at R $\sim$ 500 with the best-fit BT-Settl synthetic spectra (red).}
	\label{figure2}
\end{figure}


\begin{table*}
	\centering
	\caption{ Observation details of the sources of this study along with the derived stellar parameters. V band magnitudes are taken from the SIMBAD database. The spectral types are mostly taken from \protect\cite{lepine2013}, and \protect\cite{Jeffers2018}, except for six sources where we have derived the spectral class. These sources are marked with ($\star$). The full table for all the sources is given in Table-1 of Appendix-III in the supplementary material.}
	\begin{tabular}{llccccccc}
		\hline
		Source & Source       & Spectral   & Magnitude &  Date of     & Frame exposure time       & log g            &T$_{eff}$\\
		ID     & name         & type  &  (V-band) &  observation &   $\times$ No. of frames       &     (cm s$^{-2}$)    & (K)\\
		&  &  &           &  (UT)        &     (s)    &                   &\\
		\hline
		
		1 &   PM J03332+4615S  & M0.0  &  13.09  &  2020-12-29.602  &  300sx14  &   5.0  &  3900  \\ 
		2 &    PM J03416+5513  & M0.0  &     -    &  2021-02-01.600  &  300sx18  &   5.0  &  3800  \\ 
		3 &    PM J07151+1555  & M0.0  &  11.37  &  2021-01-30.754  &  300sx18  &   5.0  &  4000  \\ 
		- - &    - -  & - -  &  - -  &  - -  &  - -  &  - -  &  - -  \\ 
		
		\hline
	\end{tabular}
	\label{table-ObsSpecMFOSCP1}	
\end{table*}


\subsection{H$\alpha$ \& H$\beta$ Equivalent widths and their variability}
\label{SubSec-HaHbEW}

The EWs of \ha and \hb emissions are calculated as,

\begin{equation}
EW= \displaystyle \sum \left(1-\frac{F(\lambda)}{F_{c}(\lambda)}\right)\delta\lambda
\end{equation}

\noindent where $F(\lambda)$ \& $F_{c}(\lambda)$ are the line and continuum flux at wavelength $\lambda$ respectively, and $\delta\lambda$ is the pixel size in the unit of wavelength. The errors in EWs include the errors in the line and continuum fluxes and errors in wavelength calibration. Following  \cite{Hilton2010}, the spectral wavebands for the \ha nd \hb are  6557.6-6571.6 $\AA$ and 4855.7-4870.0 $\AA$ respectively. The corresponding continuum regions are  6500-6550 $\AA$ \& 6575-6625 $\AA$ for \ha and 4810-4850 $\AA$ \& 4880-4900 $\AA$ for \hb emissions. The average values of the continuum flux in these regions are chosen for the EWs estimations while summing the area under the line.

\begin{figure*}
	\centering
	\includegraphics[angle=0,width=0.95\textwidth]{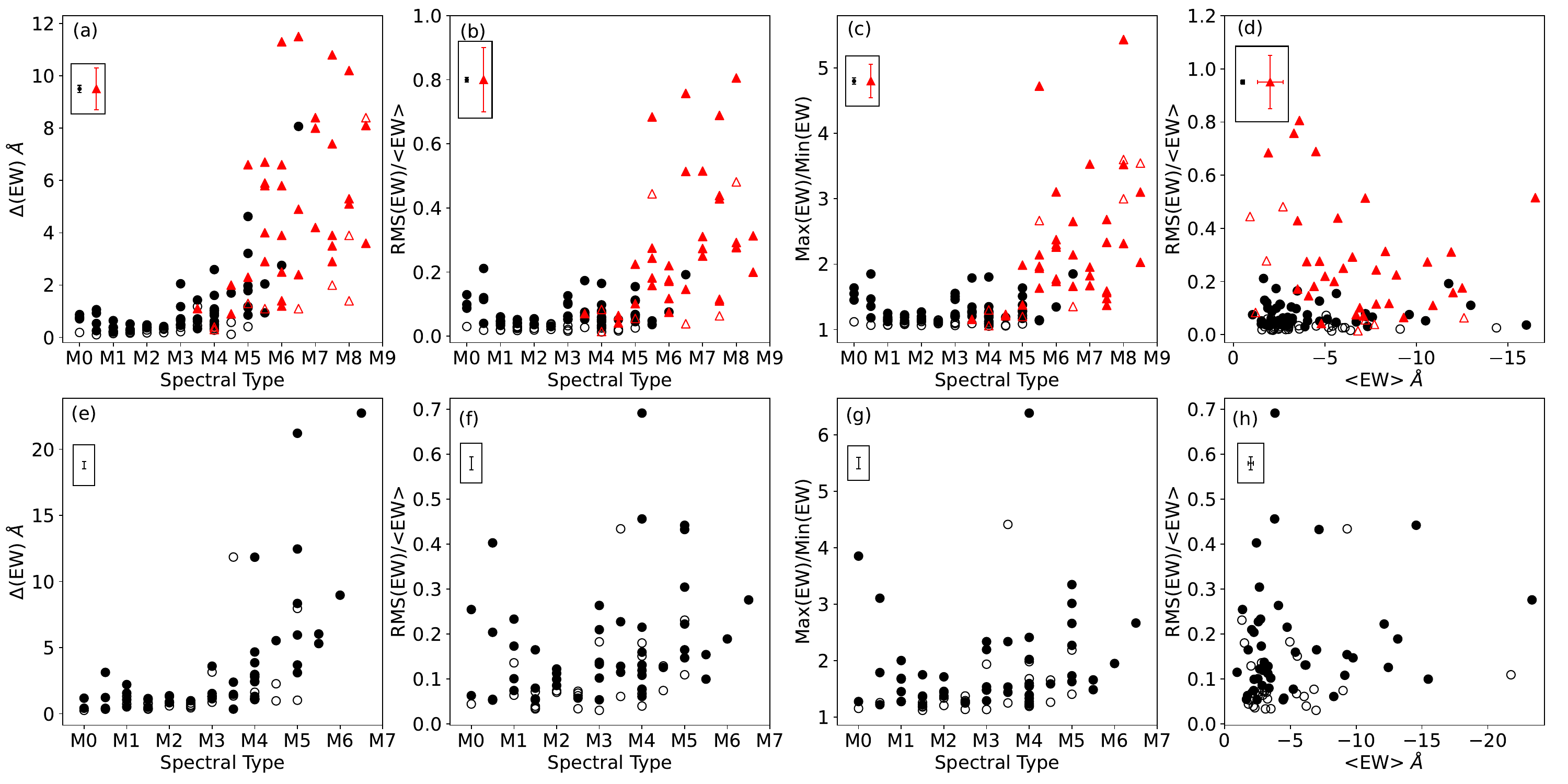}
	\caption{ Plots showing the variations of various quantities depicting the variability of the EWs of \ha and \hb emissions in M dwarfs. Panels (a), (b), and (c) (in the top row for \ha) and panels (e), (f) and (g) (in the bottom row for \hb) show the changes in  $\Delta$({\rm EW}), ${\rm RMS (EW)}/\langle {\rm EW}\rangle$, and $\rm Max(EW)/ Min(EW)$ respectively as the function of spectral type. Panels (d) and (h) show the variation of ${\rm RMS (EW)}/\langle {\rm EW}\rangle$ with respect to $\langle {\rm EW}\rangle$ for \ha and \hb respectively. Black circles represent the data points of our sample in this study. Red triangles represent the data points derived from the values given in Table-2 of \protect \cite{Lee2010}. Filled and open circles/triangles represent the objects identified with varying and non-varying \ha using the $\chi^2$ criterion ($p<$ 0.05 for variable sources). The error bars in the box on the top-left corner show the median errors of the data points. See Section ~\ref{SubSec-HaHbEW} for more details.}
\label{fig-Var_plot_HaHb}
\end{figure*}

\begin{figure*}
	\centering
	\includegraphics[angle=0,width=0.75\textwidth]{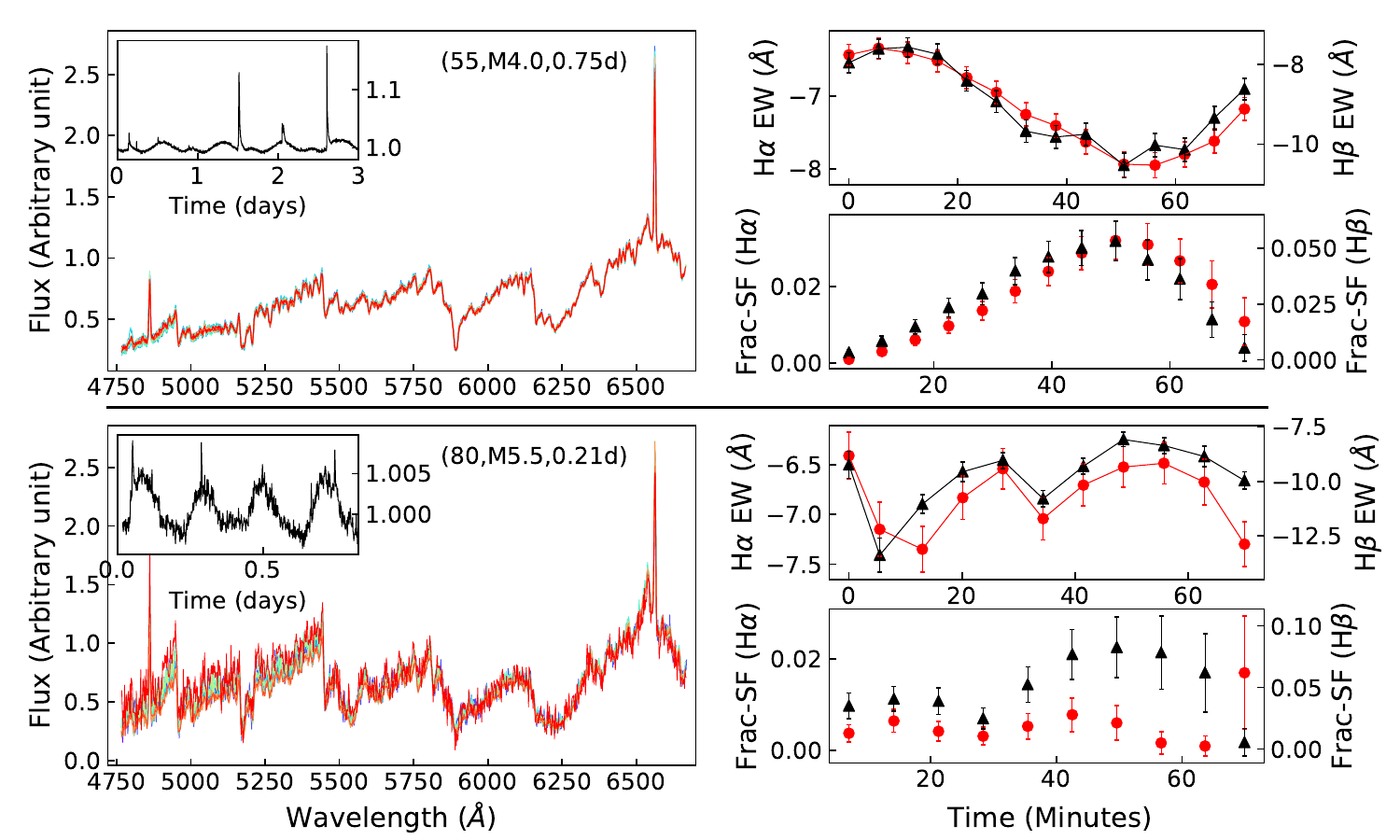}
	
	\caption{The figure shows the time-varying spectra of two sources (source ID: 55 in the top panel and Source ID: 80 in the bottom panel) along with their photometric light curves in the inset. Source ID, spectral type, and rotation period are also mentioned in each of the panels.	The corresponding upper and bottom right side panels show the time variations of the EWs of \ha / \hb and fraction structure function (SF), respectively. Data for \ha and \hb are shown in red circles and black triangles, respectively. See Section ~\ref{SubSec-HaHbEW} for discussion. Similar plots for all the sources of this study are given as Appendix-I in the supplementary material.}
	\label{fig-EW-SF}
	
\end{figure*}

As discussed in Section~\ref{sec-ObsMtAbu}, each of our sources typically has 8-18 numbers of individual spectral frames with exposure time per frame in the range of 200-600s, depending on source brightness. To quantify the emission line flux variability in this time series, a $\chi^2$ minimization is performed over the EW time series (EW light-curve) for each source. The $\chi^2$ values have been estimated by fitting a straight line at a constant EW to \ha and \hb EW light curves. The confidence of $\chi^2$ fit was determined by calculating $p$-values for given degrees of freedom, using an open-source numerical library scipy.stats.chi2\footnote{https://docs.scipy.org/doc/scipy/reference/generated/scipy.stats.chi2.html} of Python. A source is termed as a variable if its $p$-value was determined to be less than 0.05. In our sample of 83 M dwarfs, we find that 30 objects ($\sim 36\%$) show no variation in the \ha emission with the confidence level more than $95\%$ ($p$-value $<$0.05). The computed $p$-value of each source is listed in Table~\ref{table-EW1} for both \ha and \hb time series. The summary of the median ($\langle {\rm EW}\rangle$), minimum (Min), maximum (Max), and root-mean-square (RMS) values of the \ha \& \hb EWs are also given in Table~\ref{table-EW1}. 
\par
We use the same metrics to quantify the variability strength as used by \cite{Lee2010}, namely, $\Delta{\rm EW} = Max(EW) - Min(EW)$,  ${\rm RMS(EW)}/\langle {\rm EW}\rangle$ and $R({\rm EW})= \rm Max(EW)/ Min(EW)$ for both the \ha and \hb emission lines. This helps in comparing the variability observed in our sample in the range of M0-M6.5 with that of \cite{Lee2010} in the M3.5-M8.5 spectral range. Further, these quantities do not require normalization by L$_{bol}$ for comparison across spectral types \citep{Kruse2010}. In Fig.~\ref{fig-Var_plot_HaHb}, we show the variations of these quantities as a function of spectral types (panels a, b, c for \ha and panels e, f, g for \hb). As reported earlier by \cite{Lee2010,Kruse2010}, we notice a clear rising trend, which signifies the higher activity in the later types of M dwarfs. Panels (d) and (h) in Fig.~\ref{fig-Var_plot_HaHb} show the distribution of median normalized RMS values of EWs (RMS(EW) / $\langle {\rm EW}\rangle$) with respect to $\langle {\rm EW}\rangle$, for \ha and \hb respectively. Here the segregation of our data set (M0-M6.5) and \cite{Lee2010} (M3.5-M8.5) data set is more prominent. The later types of M dwarfs, though having lesser $\langle {\rm EW}\rangle$, tend to be more variable. The values of ${\rm RMS(EW)}/\langle {\rm EW}\rangle$ for our sources (M0-M6.5) are found to be below $\sim$0.2 for \ha  and below $\sim$0.5 for \hb. These trends are again discussed in Section~\ref{Sec-Discussion} along with other results.
\par
We attempted to explore the time scales of this variability, if any, using a simple construct of the fractional structure function (SF). For a given EW time series (for \ha and \hb), the fractional SF at a given time scale ($\tau$) is defined as,
\begin{equation}
SF(\tau)=\displaystyle \left\langle \left[\frac{EW(t+\tau)-EW(t)}{\rm Mean(EW)}\right]^2\right\rangle
\end{equation}

where EW(t+$\tau$) and EW(t) are two measurements of EW at the time interval of $\tau$ and Mean(EW) is the mean value of the EWs in the full-time series. The time interval values ($\tau$) have been chosen so that they will start with the minimum time interval to the maximum time intervals present in the EW series. The spectra, EW light curves, and fractional SFs for \ha and \hb are shown in Fig.~\ref{fig-EW-SF} for two of the sources of our sample. Similar plots for all the sources of our sample are given in Appendix-I in the supplementary material. Though the fractional SFs do not clearly show any characteristic time scale of variability (especially at lower times of a few minutes), they nevertheless reinforce an interesting trend noticed by \cite{Bell2012}.  

The sources which are seen to be varied at a longer time scale (e.g., source ID: 55 in the upper panel of Fig.~\ref{fig-EW-SF}) exhibit a fractional SF, which shows an increasing trend, as expected. However, conforming to the results of \cite{Bell2012}, the sources whose EWs light curves are seen to be varied at shorter time scales (e.g., source ID: 80 in the lower panel of Fig.~\ref{fig-EW-SF}) show a nearly flat distribution of fractional SF at all times. \cite{Bell2012} attributed such behavior of the highly variable sources due to the variability time scale shorter than their smallest time-separation bin of $\sim$15 minutes. We also see the same trend even at the cadence of $\sim$5 minutes.

\begin{figure*}
	\centering
	\includegraphics[angle=0,width=0.8\textwidth]{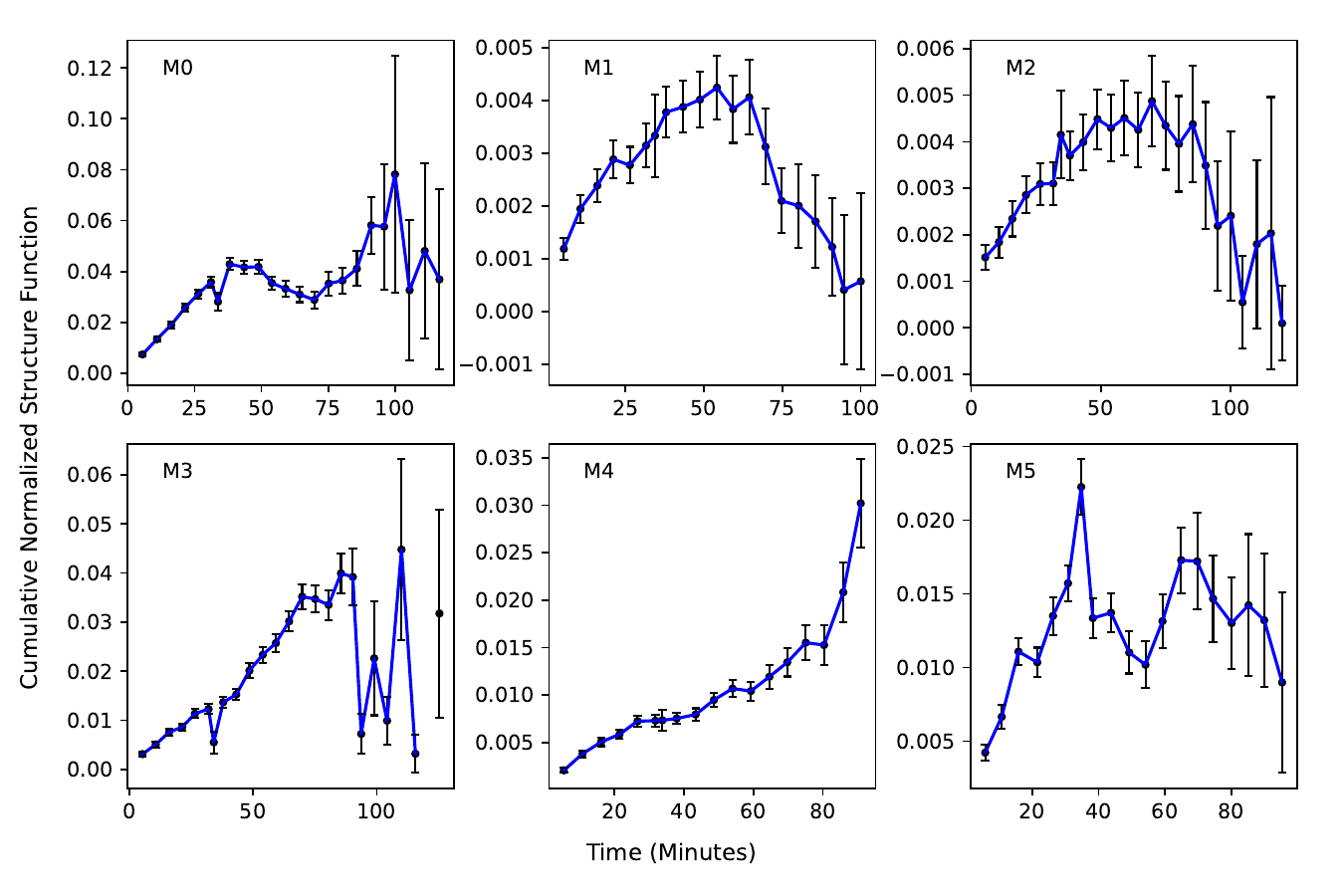}
	\caption{The figure shows the cumulative normalized structure function for various spectral classes for sources observed in this study.}
	\label{fig-SF-sptype}
\end{figure*}

In order to explore the time scale of variability for a particular spectral type (e.g., M0, M1, etc.) in our sample, we first calculated the normalized-structure function at time scale ($\tau$) as,
\begin{equation} 
Normalized~SF(\tau)=\displaystyle \left[\frac{EW(t+\tau)-EW(t)}{\rm Mean(EW(t+\tau),EW(t))}\right]^2
\end{equation}
This was done for all the EW time series (for all the sources in that particular spectral bin). Later the cumulative-normalized structure function (CNSF) at time scale $\tau$ is calculated by taking the mean of these normalized structure functions. We show the cumulative normalized structure function for different spectral bins of our sources in Fig.~\ref{fig-SF-sptype}. While a peak is seen in these plots for the early type M dwarfs (M0-M2), signifying the presence of a variability time scale ($\sim$40-60 minutes) in the \ha EW, this behavior is not very obvious in the mid-type M dwarfs (M3 and M4), thereby indicating that possibly the time scale of variability is longer than 60 minutes. For M5, it might have two peaks showing $\sim$30 minutes of variability time scale or having a significant peak at a longer time scale, thus longer monitoring is required to confirm the variability time scales. These variability time scales conform with the reported value (0.25 - 1 hour) in the other studies e.g., \cite{Lee2010, Kruse2010, Bell2012}. Though we could only observe different time scales of variability for early to mid M dwarfs, a similar result is also presented by \cite{Kruse2010}, who mentioned that the time scale of variability increases with later spectral types. We suggest that such differences in the time scales of variability could be due to different magnitudes of magnetic activity as a function of spectral type.

\subsection{\ha and \hb activity strength and flaring sources}
\label{SubSec-HaHb-Strength}

\ha or \hb activity strength is defined as the ratio of their luminosity to the bolometric luminosity \citep{West2008,Hilton2010,Lee2010,Newton2017}. The activity strength enables a better comparison of activity between stars of different masses than EW alone \citep{Reid1995b}, as it shows the importance of the line flux relative to the entire energy output of the star. We adopted the following relations given by \cite{Douglas2014} to calculate the \ha and \hb activity strength as,

\begin{equation} \label{equation-chi-factor} 
\begin{split}
\frac{L_{H\alpha}}{L_{bol}} = -\chi_{\rm H\alpha} \times (EW)_{H\alpha} \\
\frac{L_{H\beta}}{L_{bol}} =  -\chi_{\rm H\beta} \times (EW)_{H\beta}
\end{split}
\end{equation}
\begin{figure}
	\centering
	\includegraphics[angle=0,width=0.5\textwidth]{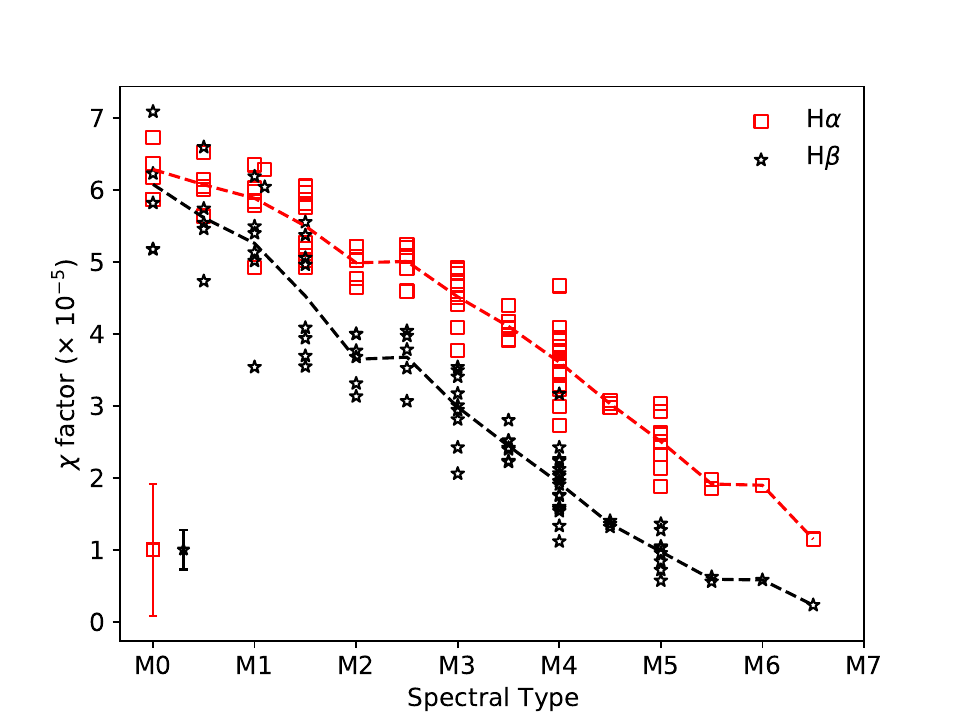}
	\caption{The derived $\chi$ factors for \ha and \hb are shown with respect to spectral types. Median error bars are shown in the bottom-left of the plot. The lines joining the mean values of $\chi$ factors for a given spectral bin are also shown for \ha and \hb.}
	\label{fig-Temp_chiha_chihb}
\end{figure}


where $EW_{H\alpha}$ and $EW_{H\beta}$ are the equivalent widths of the \ha and \hb emission lines respectively, and the $\chi$ factor for \ha and \hb are derived from photometric color $(i-J)$ \citep{Walkowicz2004, Douglas2014,West2008b}. The $\chi$ factor is defined as the ratio of the flux in the continuum near \ha to the bolometric flux \citep{Walkowicz2004}. For the sources where $i$ magnitudes are not available, we adopted the approach of \cite{Newton2017} first to calculate the M$_{Earth}$ magnitudes \citep{Dittmann2016} and later applied the relation given in Section 3.3 in \cite{Newton2017} to calculate the final $i_{48}-J$ color. Similar to \cite{Newton2017}, we have not made any additional correction between i$_{48}$ and $i$ as it would be minor. In Fig.~\ref{fig-Temp_chiha_chihb}, we show the variation of derived $\chi$ factors as a function of the spectral type where a trend similar to \cite{Newton2017} is noticed. They are also consistent with the range of $\chi$ factors given in \cite{Newton2017}. 
\par
The derived values of the means of $L_{\rm H\alpha}/L_{\rm bol}$ and $L_{\rm H\beta}/L_{\rm bol}$ are shown in Fig.~\ref{fig-Strength_HaHb} (top-right panel for \ha and bottom-right panel for \hb) with respect to their spectral type. The ratio of maximum to minimum values of these quantities (for \ha and \hb) to their mean value is shown in the top-left and bottom-left panels of Fig.~\ref{fig-Strength_HaHb}. Both of these, $L_{\rm H\alpha}/L_{\rm bol}$ and $L_{\rm H\beta}/L_{\rm bol}$, represents the activity strength of the M dwarfs.  The plots for \ha also include the values from Table-2 of \cite{Lee2010} as well. It is to be noted that \cite{Lee2010} did not provide the values of $L_{\rm H\alpha,min}/L_{\rm bol}$. Therefore we have first calculated the $\chi$ factor using $L_{\rm H\alpha,max}/L_{\rm bol}$, and the corresponding value of maximum EW (using equation~\ref{equation-chi-factor}). This $\chi$ factor was then used with the corresponding value of minimum EW (from Table-2 of \cite{Lee2010}) to calculate the value of $L_{\rm H\alpha,min}/L_{\rm bol}$.
\par
L$_{H\alpha}$/L$_{bol}$ reaches a constant value of $\sim$10$^{-3.8}$ for the M dwarfs with spectral type M0-M4 and then declines for later spectral types (later than M4), indicating the lower activity strengths for the later types. However, the variability, which can be estimated as the ratio of maximum to minimum values (of the \ha and \hb line flux for a given time series of an M dwarf), is higher for these later spectral types. This again signifies that though the strength of the activity in these late-type M dwarfs is low, they are more variable. 

\begin{figure*}
	\centering
	\includegraphics[angle=0,width=0.73\textwidth]{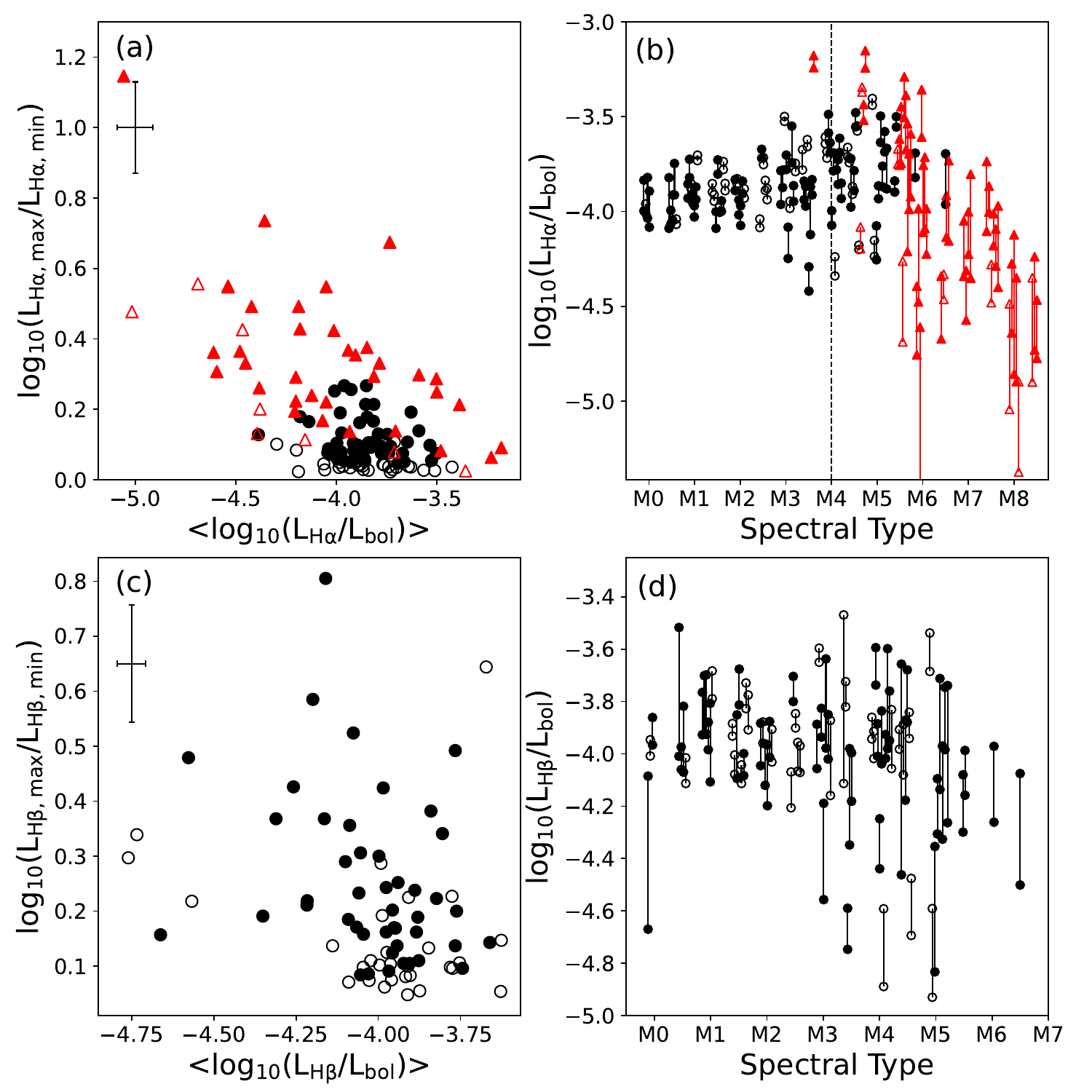}
	\caption{ The distribution of derived activity strengths ($L_{\rm H\alpha}/L_{\rm bol}$ and $L_{\rm H\beta}/L_{\rm bol}$) for \ha and \hb (top-right panel for \ha and bottom-right panel for \hb). The solid lines connect the maximum and minimum activity strength values measured for each source. The positions of objects are displaced horizontally for clarity. The vertical solid dashed line in panel (b) shows the break occur at M4, after this the activity starts to decline. Top-left and bottom-left panels show the variation of the ratio of maximum to minimum values with respect to the mean values of the activity strength \ha and \hb, respectively. Symbols have the same meaning as in Fig.~\ref{fig-Var_plot_HaHb}.}
	\label{fig-Strength_HaHb} 
\end{figure*}
\par 
Though \ha in M dwarfs is known to show variability up to 30$\%$ in the ``quiescent" phase \citep{Gizis2002,Hilton2010, Lee2010}, flaring events during the observing window could also cause additional variability in \ha and \hb measurements. Therefore, we have also checked the possibility of flaring in our sample. The short exposure spectral time series of the data in this study also allows us to quantify the possibility of flares during our observations. For this purpose, we have utilized the method proposed by \cite{Hilton2010} by determining the ``flaring line index" (FLI). FLI for the \ha and \hb lines are defined as \citep{Hilton2010},
\begin{equation}
FLI= \bar{l}/ \sigma 
\end{equation}
\noindent where $\bar{l}$ is the mean value of the continuum subtracted flux in the line region, and $\sigma$ is the standard deviation of the continuum. FLIs are useful where emission-line strengths are weak and/or continuums are noisy. The statistics of FLIs of a given spectral time series would then be used to decide if a flare occurred in the observing window, based on the following criteria by \cite{Hilton2010}:\\ 

\noindent (1) For strong emission line sources (i.e, mean FLI values for both \ha and \hb emission of a time series are $>$ 3), if the maximum(FLI) and minimum(FLI) are differed by more than 30\%, they are characterized as a flaring source during the time series observations. \\
(2) For weak emission line sources (i.e, mean FLI values for both \ha and \hb emission of a time series are $<$ 3), if the maximum(FLI) - minimum(FLI) $>$ 3, they were classified as flaring sources. \\
(3) For those sources which could not be categorized as strong/weak emission line sources (i.e, mean FLI value $>$3 for \ha but $<$ 3 for \hb), if both the above two conditions were satisfied, they were classified as flaring sources. \\

Thus, after applying the above criteria, we found that out of 75 objects that showed both the \ha and \hb emissions, 53 objects were found to be in a flaring state at the time of observations. Out of these 53 objects, 38 were earlier classified as variable sources based on the $\chi^2$ minimization method as discussed in Section~\ref{SubSec-HaHbEW}.

 Thus, it appears that the flaring events could be a major cause of short-term variability seen in the EW light curves. Table~\ref{table-TEES1} shows the flaring status of all 75 objects along with their mean FLI values for \ha and \hb.  \cite{Hilton2010} noticed that the short-term variability in chromospheric emission could be due to low-level flaring events. In a recent study by \cite{Medina2022}, the author suggested that the flares are the most suitable mechanism to describe \ha variability on active mid-to-late M dwarfs. They have also mentioned that this scenario is not suitable for inactive early M dwarfs, where this chromospheric emission variability comes from the roughly constant emission from star spots or plages rotating into and out of view on inactive early M-dwarfs. However, there are very few studies that tried to explore the short-term variations and hence, a better time resolution could very well be a key to establishing such a hypothesis. 
\begin{table*}
	\centering
	
	\caption{The derived variability parameters for \ha and \hb emission lines for the sources of this study, along with the activity strengths ( log$_{10}$(L$_{H\alpha}$/L$_{bol}$) and log$_{10}$(L$_{H\beta}$/L$_{bol}$)). The $p$-values are determined from the $\chi^{2}$ minimisation of EW light curves. See Section~\ref{SubSec-HaHbEW} for details. The sources which are characterized as variable are marked with ($\star$). The full table for all the sources is given in Table-2 of Appendix-III in the supplementary material.}
	\resizebox{\textwidth}{!}{  
		\begin{tabular}{llcccccccc}
			\hline
			Source & Source   & emission   & Median  & Minimum   & Maximum &  $\Delta$ & RMS  & Mean  & P-value\\
			ID &     name     & line       & \ha EW & \ha EW   &  \ha EW &  \ha EW   & \ha EW & log$_{10}$(L$_{H\alpha}$/L$_{bol}$) & \ha \\
			& &    & \hb EW & \hb EW   &  \hb EW &  \hb EW   & \hb EW & log$_{10}$(L$_{H\beta}$/L$_{bol}$) & \hb \\	
			\hline
			1 & PM J03332+4615S$^\star$  & \ha &  -2.333 $\pm$ 0.065  &  -1.717 $\pm$ 0.062  &  -2.492 $\pm$ 0.061  &  0.775 $\pm$ 0.087  &  0.232 $\pm$  0.022  &    -3.883  &  0.0000  \\ 
			   &   & \hb &  -1.372 $\pm$ 0.085  &  -0.413 $\pm$ 0.091  &  -1.590 $\pm$ 0.086  &  1.177 $\pm$ 0.125  &  0.349 $\pm$  0.031  &    -4.199  &  0.0000  \\ 
			2 &   PM J03416+5513  & \ha &  -1.688 $\pm$ 0.047  &  -1.587 $\pm$ 0.045  &  -1.775 $\pm$ 0.052  &  0.188 $\pm$ 0.068  &  0.051 $\pm$  0.012  &    -3.984  &  0.2298  \\ 
			  &   & \hb &  -1.818 $\pm$ 0.085  &  -1.690 $\pm$ 0.076  &  -1.947 $\pm$ 0.087  &  0.258 $\pm$ 0.116  &  0.080 $\pm$  0.020  &    -3.981  &  0.4134  \\  
			- - &    - -  & - -  &  - -  &  - -  &  - -  &  - -  & - -  &  - -  & - -  \\
			\hline
		\end{tabular}
	}
	\label{table-EW1}	
\end{table*}

\section{Results from Photometry: Rotation Period and Star-spot Filling Factor}
\label{sec-phot_result}
In the last decade, TESS and Kepler/K2 missions \citep{Caldwell2010, Koch2010, Howell2014, Ricker2015} have provided unprecedented cadence and photometric precision for a variety of stellar sources. Such high cadence data is complementary to our spectroscopy analysis. Thus, we have utilized these photometric light curves to determine the rotation periods of the objects in our sample. While for this study, we wished to explore the period dependence of various activity proxies (\ha EW, L$_{H\alpha}$/L$_{bol}$, etc.). Some of these findings are discussed below.
\par 
70 of the 83 M dwarfs in our sample were observed by TESS in 2014-2021 in various sectors. Data reduction pipelines from the Science Processing Operations Center (SPOC: \cite{Jenkins2016}) and the Quick Look Pipeline (QLP: \cite{Huang2020}) are used to derive the light curves of these sources, where the Simple Aperture Photometry (SAP) flux is used. For most sources, data with 2 minutes cadence are available and used in this study. In some cases, where this high cadence data were not available, 30 minutes cadence data were used. Of the remaining 13 sources, where TESS light curves were not available, we could find the photometric light curves of 5 of the sources, with 30 minutes cadence, in the Kepler/K2 data archive. The details are given in Table~\ref{table-TEES1}. TESS and Kepler/K2 light curves are also used to determine the rotation periods of the objects from the \cite{Lee2010}. Light curves of 31 (of total 43) objects were found in these archives. The details are given in Table~\ref{table-TESS_Lee}.

\subsection{Rotation Period}

The rotation periods (P$_{rot}$) are measured by quantifying the periodic brightness variations in the light curve, which are caused by the starspots on the surface of the objects. These rotation periods have been determined by the Generalized Lomb-Scargle periodogram as discussed in \cite{Zechmeister2009}. The period of the maximum power was further cross-checked with the phase folded light curves by visual inspection. Out of 106 sources where the light curves were available, the rotation periods of 82 objects were determined with the above method.

 \begin{figure}
	\centering
	\includegraphics[angle=0,width=0.5\textwidth]{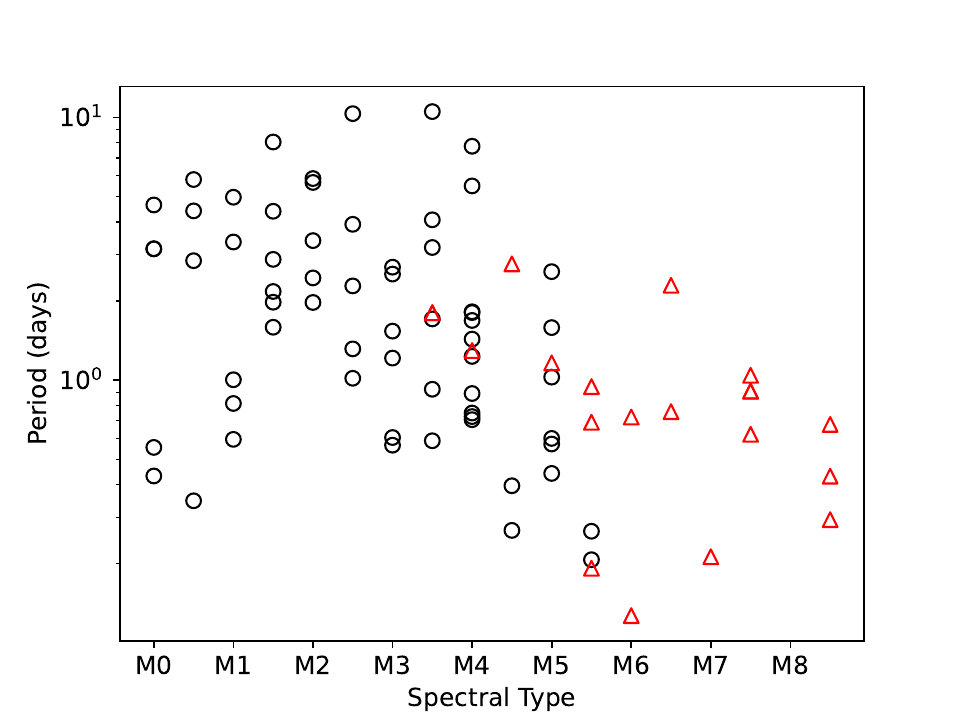}
	\caption{Distribution of the derived rotation periods of M dwarfs with the spectral type. Sources of this study and from \protect \cite{Lee2010} are shown in black circles and in red triangles, respectively.}
	\label{fig-RotPeriod-SpecType}
\end{figure}

\begin{figure*}
	\centering
	\includegraphics[angle=0,width=\textwidth]{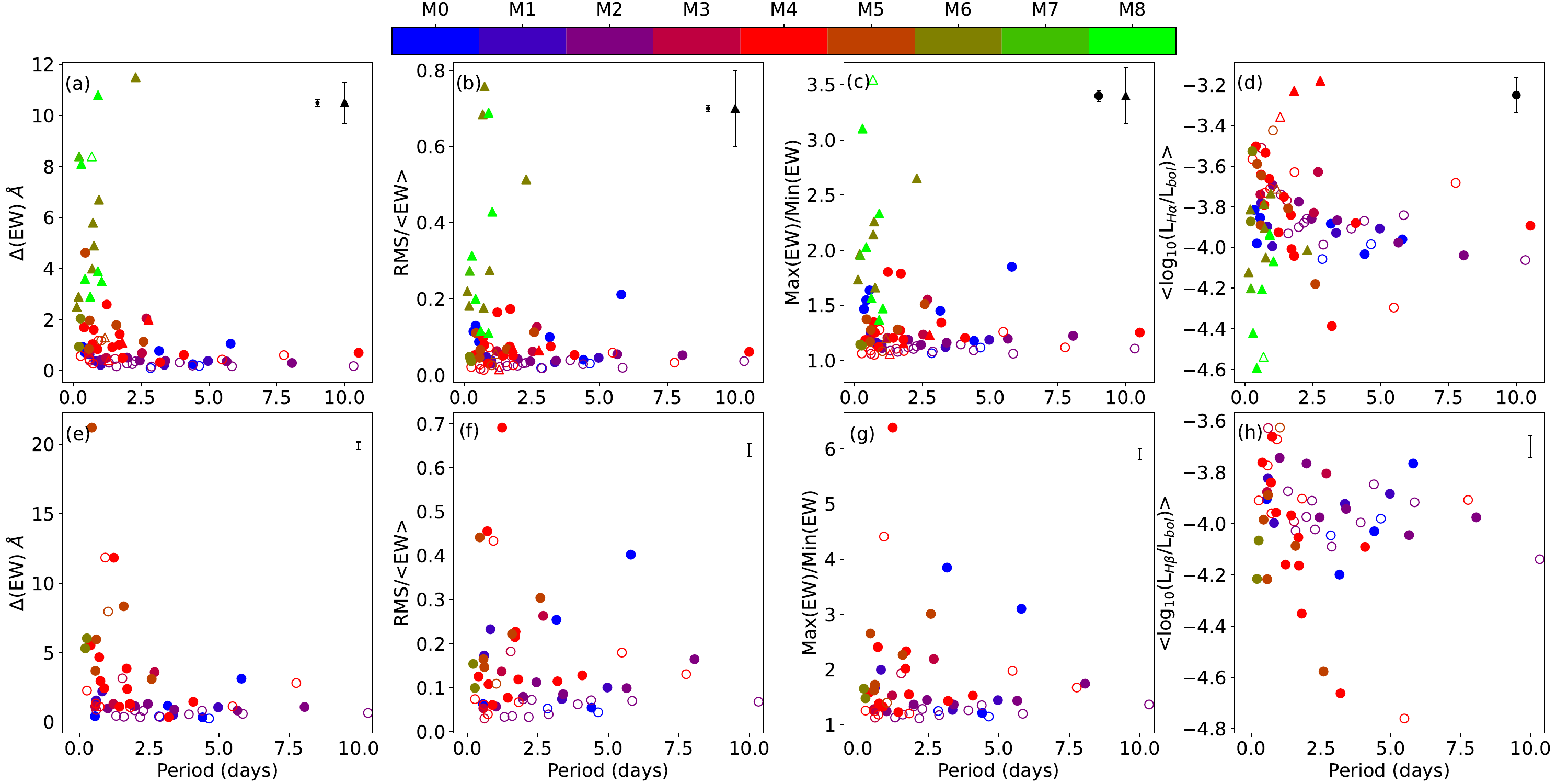}
	\caption{Figure shows the distribution of $\Delta({\rm EW})$, ${\rm RMS(EW)}/\langle {\rm EW}\rangle$, $R({\rm EW})= \rm Max(EW)/ Min(EW)$ and mean activity strengths with their rotational periods. Top panels are for \ha and bottom panels are for \hb emission. The median errorbars of the data points are shown on the top-right corner. Symbols have the same meaning as in Fig.~\ref{fig-Var_plot_HaHb}}
	\label{fig-variability-wrt-period-HaHb}
\end{figure*}

\begin{figure*}
	\centering
	\includegraphics[angle=0,width=0.9\textwidth]{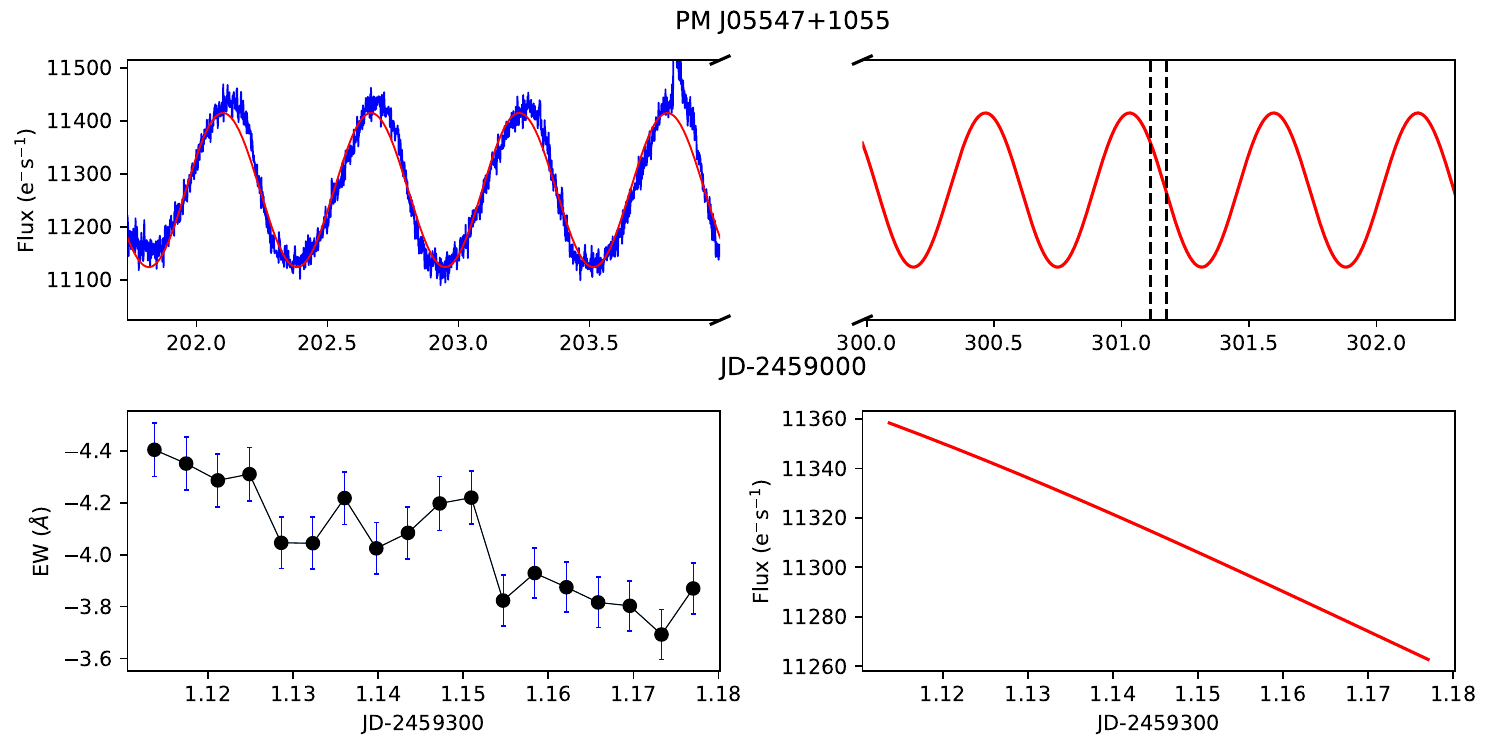}
	\caption{The top panel shows the extrapolated sinusoidal fit (red) with the photometric light curve (blue). The vertical line shows the start and end of our monitoring spectroscopic observations. The bottom left and right panels show the \ha equivalent width light curve and corresponding extrapolated photometric light curve, respectively, showing a correlation between the star's brightness and \ha EW.}
	\label{fig-Period-track}
\end{figure*}

\begin{figure*}
	\centering
	\includegraphics[angle=0,width=0.9\textwidth]{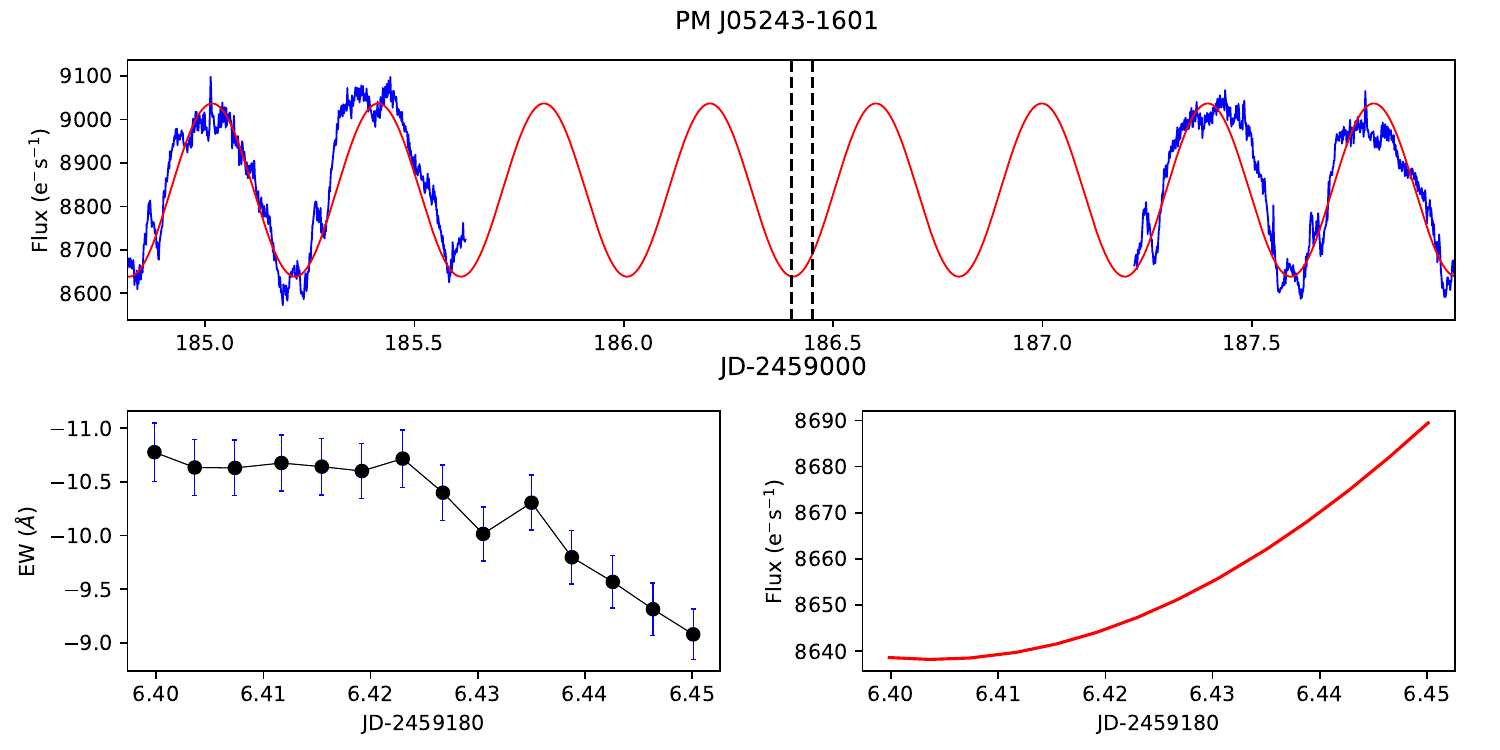}
	\caption{The figure description is the same as Fig.~\ref{fig-Period-track} but shows an anti-correlation between the star's brightness and \ha EW.}
	\label{fig-Period-track2}
\end{figure*}

 \par
The rotation periods of all the objects are listed in Table~\ref{table-TEES1} and Table~\ref{table-TESS_Lee}, respectively. They are found to be in the range of $\sim$0.2-10 days. These periods are plotted against the spectral type in Fig.~\ref{fig-RotPeriod-SpecType}. The derived periods conformed to the general trend seen in the other studies \citep{West2015, Newton2017, Jeffers2018}, wherein the rotation periods were found to be shorter for later spectral types. The distribution of the variability indicators, namely, $\Delta${\rm EW} , ${\rm RMS (EW)}/\langle {\rm EW}\rangle$, and $\rm Max(EW)/ Min(EW)$ as well as activity strengths ($L_{\rm H\alpha}/L_{\rm bol}$ and $L_{\rm H\beta}/L_{\rm bol}$) (see Sections~\ref{SubSec-HaHbEW} and ~\ref{SubSec-HaHb-Strength}) with respect to rotation periods are shown in Fig.~\ref{fig-variability-wrt-period-HaHb}.
\par
Various magnetic activity indicators have been used in the past to explore the relationship between the magnetic field strength and stellar rotation \citep{Douglas2014,West2008,West2015, Newton2017,Jeffers2018}. Similar to \cite{West2015}, \cite{Jeffers2018}, and \cite{Duvvuri2023}, we also find that M dwarfs with longer periods show less variability. \cite{Duvvuri2023} suggested that these trends may be a function of the total surface area of active regions, the depth of the chromosphere subject to variable energy deposition via microflares or Alfvén waves, magnetic field topology, etc. It has also been known that in M dwarfs, there is a clear decrease in the strength of activity with increasing rotation period \citep{West2015, Jeffers2018}. However, an interesting behaviour can be noticed in the panels $a$, $b$, $c$ (for \ha) and $e$, $f$, $g$ (for \hb), when we consider the spectral types as well. Here the short-term variability indicators display a very clear increase for the faster rotating M dwarfs with a rotation period $<$ 2 days, and most of these objects belong to the later types (M4-M8) as seen in Fig.~\ref{fig-RotPeriod-SpecType}. Here we intend to show the variation of the rotation period as a function of the spectral type we have derived for our sample. Our analysis is consistent with the trends seen in other studies such as \cite{Jeffers2018}. Thus such high variability in late-type fast rotators M dwarfs reflects an overall shift in their magnetic field, leading to less uniform spatial and temporal dissipation as suggested by \cite{Kruse2010}.

\par 
Another important parameter is the Rossby number (R${_0}$), which is the ratio of the rotation period to the convective overturn timescale. It is used to compare activity strengths across mass and rotation period ranges. We have used the empirical calibration from \cite{Wright2011} (Eq.~10, page-11) to determine convective overturn timescales. Our samples show a saturated relationship between activity strengths and R${_0}$ (see Fig.~\ref{fig-Rossby_number}). Since the Rossby number for our samples ranges from 0.0003-0.2, the observed saturation is expected. A power-law decay in $L_{\rm H\alpha}/L_{\rm bol}$ with increasing R${_0}$ ($>$0.2) could not be seen, which breaks near R${_0}$$\sim$0.2.

\begin{figure*}
	\centering
	\includegraphics[angle=0,width=0.7\textwidth]{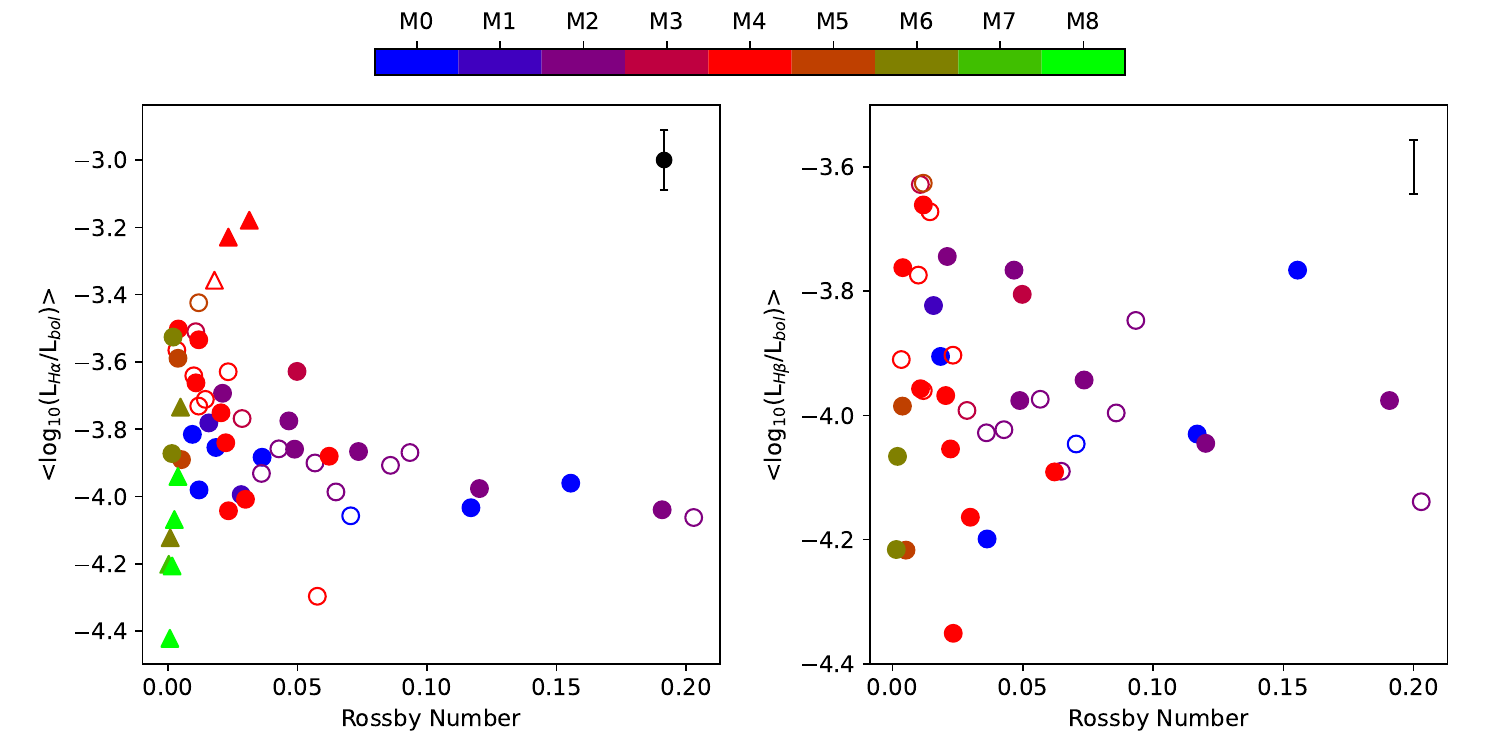}
	\caption{ The distribution of derived activity strengths $L_{\rm H\alpha}/L_{\rm bol}$ (left panel) and $L_{\rm H\beta}/L_{\rm bol}$ (right panel) with respect to Rossby number.  The median errorbars of the data points are shown on the top-right corner. Symbols have the same meaning as in Fig.~\ref{fig-Var_plot_HaHb}.}
	\label{fig-Rossby_number}
\end{figure*}

\par 

There are some studies \citep{Radick1998, Hall2009, Buccino2011, Fuhrmeister2019, Shoefer2022, Medina2022, Duvvuri2023}, which estimate the rotation period using chromospheric activity indicators (\ha, Ca II IRT, etc.) or using photospheric activity indicators (TiO 7050 absorption band, etc.). \cite{Fuhrmeister2019} emphasized that \ha and the Ca II IRT lines are well suited for period searches only if the star is dominated by a periodic pattern of \ha or Ca II plages and not dominated by micro-flaring and another intrinsic variability. These studies suggest that young, active stars become fainter as their chromospheric emission (e.g., \ha) increases, whereas older, less active stars become brighter as their chromospheric emission increases. These studies also interpreted the cause of this behavior as the long-term variability of young stars is spot dominated, whereas, for older stars, it is faculae dominated. Therefore, it is instructive to explore such behavior in spite of having a relatively shorter time series. The typical rotation period of the stars in our sample is in the range of 0.2-10 days, while our spectral time series for a single source is around 0.7-2.3 hours. Thus, the observed fraction of the rotation period of the stars in our sample is very small, ranging from 0.4 to 24\% of the period. We have, nevertheless, tried to map the derived EWs of \ha with respect to the stellar rotation phase. 

\par

To determine the phase of the object during the observations of the spectral time series, we have extrapolated the photometric light curve to our observing window by fitting a sinusoidal function, assuming the photometric light curve modulation does not vary significantly. We could do this for 49 of the sources. The analyses for two of the sources are shown in Fig.~\ref{fig-Period-track} and Fig.~\ref{fig-Period-track2}. While the plots for all the sources are given in Appendix-II in the supplementary material. We do notice that five sources show (visually) an ‘apparent’ correlation of \ha EW with the phase of the photometric light curve (i.e., \ha EW increases with increases in the photometric brightness), and another nine sources show an ‘apparent’ anti-correlation. However, such ‘apparent’ relations are noticed only visually, and therefore we do not attempt to draw any conclusion here. A longer time series will indeed be very useful to say anything conclusively regarding the presence of such plausible correlations. We shall attempt to further obtain a longer time series for these sources for such purposes.

\begin{table*}
	\centering
	\caption{Derived rotation periods and star-spot filling factors of the objects of this study using TESS and Kepler/K2 light curves. The rotation periods from the literature survey are also mentioned with references and remark (if any). The observing details from TESS and Kepler/K2 archives and computed mean values of FLI for \ha and \hb emissions from the spectral time series are also mentioned. The flaring sources (as per FLI criteria) during the spectroscopy observations are marked with ($\star$) in the source-name column. See Section~\ref{sec-phot_result} for discussion. In the third column (mission/year/author), the notation `T' represents TESS sector and a `K’ represents K2 campaign. The full table for all the sources is given in Table-3 of Appendix-III in the supplementary material.}
\resizebox{\textwidth}{!}{  
	\begin{tabular}{llcccccccccc}
		
		\hline
		Source     & Source     & Mission/Year/ Author & Exposure  & Rotation        & Rotation Period from      & Ref. & T$_{spot}$    & Filling factor & Mean FLI & Mean FLI  \\
		ID            & name        &                               & time (s)    & period (days) & literature$^b$ (days)            &        &  (K)               &     \%          & \ha         &    \hb       \\
		\hline
		
		1	&	PM J03332+4615S$^\star$ 	&	  T-18/2019/QLP 	& 1800  &	3.160	&	-			&	-	&	3192	&	36	&	 10.40 $\pm$ 1.72  	&	   2.65 $\pm$ 0.95  \\
			2	&	PM J03416+5513$^\star$  	&	  T-19/2019/SPOC	&	120	&	4.641	&	8$^1$		&	4	&	3145	&	6.2	&	  6.97 $\pm$ 0.62  	&	   4.50 $\pm$ 0.45  	\\
	3	&	PM J07151+1555$^\star$  	&	  T-33/2020/SPOC	&	120	&	0.554	&	0.555		&	6	&	3239	&	9.6	&	 12.05 $\pm$ 1.83  	&	   4.67 $\pm$ 0.42  \\		
		- - &    - -  & - -  &  - -  &  - -  & - - &    - -  & - -  &  - -  &  - -  & - - \\
		\hline
	\end{tabular}
}
	\label{table-TEES1}
	\begin{list}{}{}
		\item a: References- 1. \cite{Messina2017}, 2. \cite{Houdebine2016}, 3. \cite{Vidotto2014},  4. \cite{Magaudda2020}, 5. \cite{Newton2017}, 6. \cite{Kiraga2012}, 7. \cite{Schoefer2019}, 8. \cite{Stelzer2022}, 9. \cite{Rodriguez2020}, 10. \cite{Raetz2020}, 11. \cite{Newton2016}, 12. \cite{Gunther2020}, 13. \cite{Newton2018}, 14. \cite{Houdebine2017}.
		\item b: Sources having different rotation periods from this study, computed with the following method: 1. This was estimated by the empirical relation between Chromospheric activity and rotation period, 2. This was estimated by using All Sky Automated Survey (ASAS) photometry data (having low cadence; mostly one data point per day), 3. This was estimated by Hungarian-made Automated Telescope Network (HATNet) survey photometry data (they studied 1568 stars), 4. This was estimated with the help of v-sin$i$ and the radius of the star, where the star's radius was calculated using empirical relation, 5. \cite{Messina2016} estimated this rotation period using their own photometry observation of this previously known visual binary stars, 6. This was estimated by using photometry data from MEarth (having cadence of 30 minutes; they studied 387 stars).
	\end{list}	
\end{table*}

\begin{table*}
	\centering
	\caption{The derived rotation periods and star-spot filling factors for 31 sources of \protect\cite{Lee2010}. The $\teff$ values for these sources are estimated from $\teff$ versus spectral type relation given in \protect \cite{Rajpurohit2013}. See Section~\ref{sec-phot_result} for discussion. The full table for all the sources is given in Table-4 of Appendix-III in the supplementary material.}
	\resizebox{\textwidth}{!}{  
	\begin{tabular}{lcccccccc}
		\hline
		
		 Source     & Mission/Year/ Author & Exposure  & Rotation        & Rotation Period from      & Ref. & T$_{eff}$  & T$_{spot}$    & Filling factor   \\
		 name        &                               & time (s)    & period (days) & literature (days)            &        &  (K)          &  (K)                &     \%           \\
		
		\hline
		
		G99-049					&	T-33/2020/SPOC      	&	120	&	1.805	&	1.809	&	5	&	3100	&	2792	&	1.2		\\
		LHS1723					&	T-32/2020/SPOC      	&	20	&	-		&	88.5	&	5	&	3100	&	2792	&	0.5		\\
		L449-1					&	T-32/2020/SPOC      	&	120	&	1.296	&	-		&	-	&	3100	&	2792	&	2		\\

		- - &    - -  & - -  &  - -  &  - -  & - - &    - -  & - -  &  - - \\
		\hline
	\end{tabular}
	}
	\label{table-TESS_Lee}	
\end{table*}

\subsection{Filling Factor}
\label{SebSec-Fillingfactor}

\par 
The light curves from TESS and Kepler/K2 are also used to determine the star-spot filling factors. We have computed these values for the objects in this study as well as from \cite{Lee2010}. The star-spots on the photosphere are the regions where the magnetic flux is much stronger, and most of the stellar flares occur. The star-spot filing factor gives the fractional area covered by star-spots (A$_{spot}$/A$_{star}$). The filling factors are computed by the following relations \citep{Jackson2013,Maehara2017,Guenther2021} :\\ 

\begin{equation}
$$\dfrac{A_{spot}}{A_{star}}=\left(\dfrac{\Delta \rm F}{\rm F}\right)_{spot}\left[1-\left(\dfrac{T_{spot}}{T_*}\right)^4\right]^{-1}$$
\end{equation}

\begin{figure*}
	\centering
	\includegraphics[angle=0,width=\textwidth]{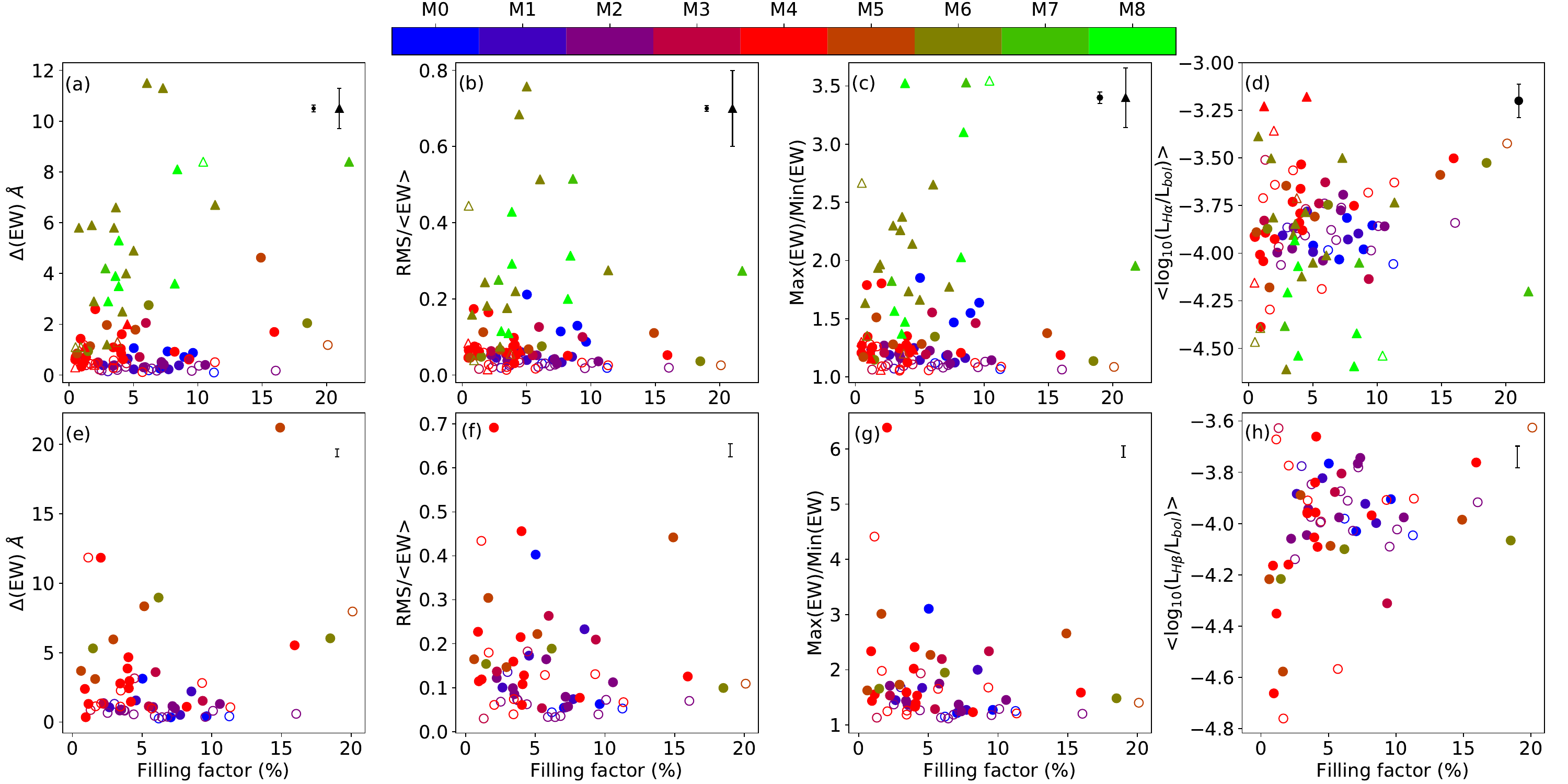}
	\caption{Distribution of variability indicators (same as Fig.~\ref{fig-variability-wrt-period-HaHb}) but with respect to star-spot filling factors.}	
	\label{fig-variability-wrt-fill-factor-HaHb}
\end{figure*}

\noindent where A$_{star}$ is the area of the stellar disk, A$_{spot}$ is the area of the spots on the stellar disk, and T$_*$ and T$_{spot}$ are the temperature of the star and the star-spot respectively. $\Delta$F/F$_{spot}$ is the brightness variation amplitude of the rotation modulation caused by the star-spots. The temperature difference between the photosphere and starspots is a function of the photospheric temperature of the stars \citep{Berdyugina2005}. Thus, T$_{spot}$ can be estimated with the following relation \citep{Berdyugina2005, Maehara2017, Notsu2019, Guenther2021}:

\begin{equation}
$$T_{spot}=T_*-3.58\times10^{-5}T_*^2-0.249\times T_*+808 [K]$$ 
\end{equation}

For the sources in this study, we have used $\teff$ as determined in Section~\ref{sec-spec_result} whereas for the sources of \cite{Lee2010} $\teff$ are determined using the spectral type - $\teff$ relation of \cite{Rajpurohit2013} (see Fig.5 therein). 
\par
The derived filling factors are given in Table~\ref{table-TEES1}. We note that these filling factors are the lower limits as many active stars could have polar spots \citep{Guenther2021}. For our samples, the spot temperatures and filling factors are mostly found between $\sim$2700-3200 K and $\sim$0.5-20.0\%, respectively. 
While for the sources of \cite{Lee2010}, these values are mostly found to be in the range of $\sim$2460-2800 K and $\sim$0.5-21.8\%, respectively.
\par 
We have also explored the variations in \ha and \hb variability indicators with respect to the derived filling factors. Fig.~\ref{fig-variability-wrt-fill-factor-HaHb} show these plots similar to Fig.\ref{fig-variability-wrt-period-HaHb} (for rotations). It is postulated \citep{Bell2012}  that stars having large filling factors would have high activity strength and lesser variability and vice versa. This trend is noticeable in panels  (b) and (f) for \ha and \hb respectively, i.e., RMS(EW) / $<$EW$>$ show a downward trend for higher filling factors, i.e., high activity stars (large $<$EW$>$) which are less variable (low RMS(EW)) tend to have high filing factors.

\section{Ages}
\label{sec-age}
As stars are gravitationally bound, they orbit the centre of the Galaxy. While interacting with various giant molecular clouds and other passing stars, their orbit gets altered via kinematic kick due to their interaction with other objects, causing the stars to separate from the plane of the Galaxy as they age \citep{Kiman2019}. The ages of stars in the StarHorse project are determined using a Bayesian approach that incorporates geometric, age, and metallicity priors for the major components of the Galaxy \citep{Queiroz2018, Anders2019}). To compute these ages, a likelihood is calculated by comparing observed parameters with the PAdova and TRiestre Stellar Evolution Code (PARSEC; \cite{Bressan2012}). In the present study, the observable quantities used include surface temperature and gravity values derived within this paper, as well as magnitudes obtained from the 2MASS survey \citep{Cutri2003} and parallaxes and magnitudes from Gaia Data Release 3 \citep{Gaia2022}. However, it is important to note that the ages estimated by StarHorse for this particular sample may exhibit significant uncertainties due to the absence of metallicity measurements and the increased complexity of isochrone fitting outside the subgiant range (see \cite{Queiroz2023} for further details). Table-\ref{table-Age-Ours1} gives the details of the parameters used to estimate the age with derived ages.
\par


\begin{table*}
	\centering
	\caption{Details presented in this table are used to determine the age (last column) of our sources. The coordinates RA, DEC and J,H,K magnitudes are taken from SIMBAD and all other parameters taken from Gaia Data Release 3 \citep{Gaia2022}. The full table for all the sources is given in Table-5 (our sources) and Table-6 (\protect\cite{Lee2010} sources) of Appendix-III in the supplementary material.}
	\resizebox{\textwidth}{!}{ 
	\begin{tabular}{cccccccccccccc}
		
		\hline
		Source     & RA    		  &         DEC           & J & H & K & $g$ & $bp$ & $rp$ &Proper motion in RA  & Proper motion in DEC    & Radial velocity &  Parallaxes  &    Age    \\
		ID           & (hh:mm:ss)        &     (dd:mm:ss)    &   &   &   &  &  &   & (mas year$^{-1}$)    & (mas year$^{-1}$)       & (km s$^{-1}$) &  (mas)        &  (Gyr)       \\
		\hline
		
		1	&	03 33 14.04	&	+46 15 18.97	&	8.382	&	7.770	&	7.592	&	10.497	&	11.305	&	9.590	&	68.9023		$\pm$	0.0146	&	-172.5847	$\pm$	0.0146	&	-14.2682$\pm$	8.3249	&	27.4809	$\pm$	0.0158		&	3.250	$\pm$	1.475	\\
2	&	03 41 37.27	&	+55 13 06.83	&	8.347	&	7.649	&	7.499	&	10.557	&	11.469	&	9.607	&	96.7821		$\pm$	0.0149	&	-116.6387	$\pm$	0.0155	&	-4.6260	$\pm$	0.3481	&	27.9199	$\pm$	0.0145		&	6.970	$\pm$	4.216	\\
3	&	07 15 08.78	&	+15 55 44.95	&	8.741	&	8.144	&	7.973	&	10.972	&	11.686	&	10.007	&	209.7643	$\pm$	0.0247	&	-169.0688	$\pm$	0.0206	&	21.2126	$\pm$	29.2719	&	18.9026	$\pm$	0.0207		&	 		-				\\
		- -           &  - -       &    - -    &  - -  &   &  - -  &  - - &  - -  &  - -  &  - -    &  - -      &  - - &   - -     &   - -       \\

		\hline
	\end{tabular}
	}
	\label{table-Age-Ours1}
\end{table*}


\begin{figure*}
	\centering
	\includegraphics[angle=0,width=0.99\textwidth]{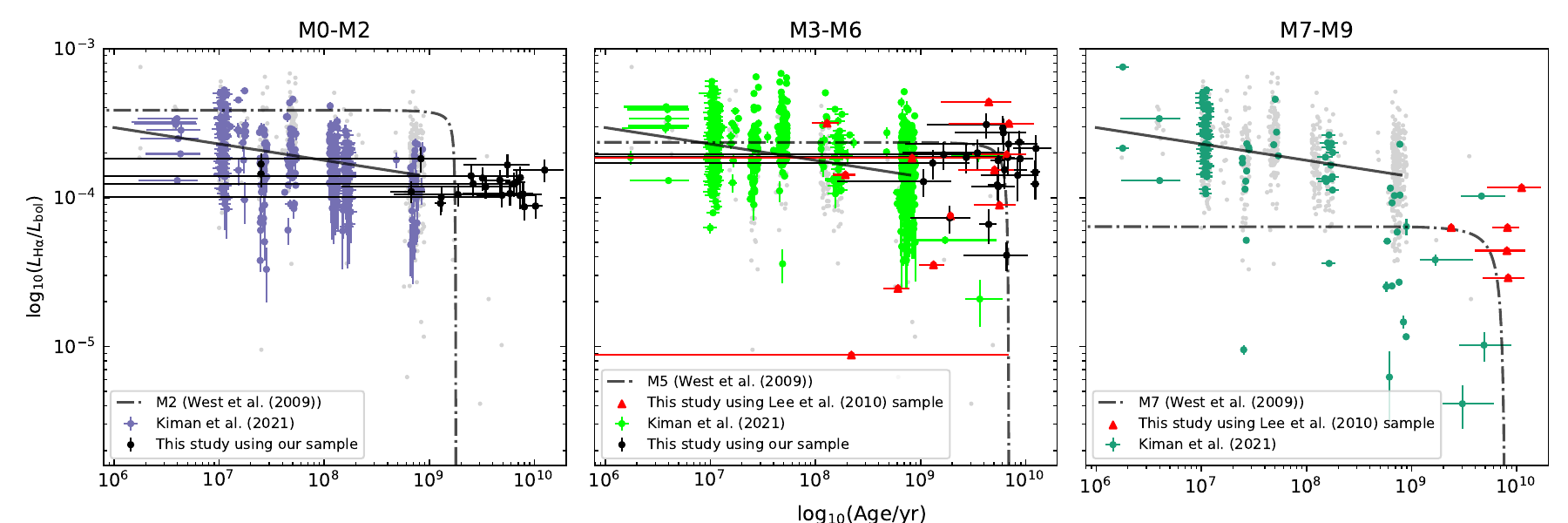}
	\caption{The figure shows the derived values of activity strength ($L_{\rm H\alpha}/L_{\rm bol}$) and age for the source in this study (black circles and red triangles) against similar data provided by \protect\cite{Kiman2021}  (see figure-10 of their article). The empirical relations provided by \protect\cite{West2009} (dot-dashed line) and \protect\cite{Kiman2021} (solid line) are also plotted. These plots are reproduced from the data presented in \protect\cite{Kiman2021} by utilising the plotting script provided by them. The background grey data points in each panel represent the complete sample of \protect\cite{Kiman2021}.}
	\label{fig-age-activity}
\end{figure*}

The \ha luminosity ($L_{\rm H\alpha}/L_{\rm bol}$) can be used as the age indicator since the young stars tend to have higher magnetic activity, whereas the old stars are less active or inactive \citep{West2008, Newton2017}. However, these relationships are not sustained because we currently do not have the data to substantiate them. Although it could be the case that fully convective stars do not follow this particular relationship. The possibilities for any such relations, in the case of M dwarfs, have been investigated in various recent studies \citep{Riedel2017, West2008, Newton2017, Kiman2021}. 

\par

In Fig.~\ref{fig-age-activity}, we have shown the derived age of stars in our sample against the similar data provided by \cite{Kiman2021} where the activity strength ($L_{\rm H\alpha}/L_{\rm bol}$) is attempted to be correlated with the age of the stars. To compare our values with them, we have also divided our samples into spectral type bins (M0-M2, M3-M6, and M7-M9). The age-activity strength values for data of our study seem to be matched with the relation given by \cite{Kiman2021} for early to mid-type of M dwarfs, though they differ from the relation provided by \cite{West2009}. This difference has also been noted by \cite{Kiman2021} for their data set, who remarks that it is likely that \cite{West2009} has overestimated the activity strength of their sample. We also noticed that the age of early M dwarfs (M0-M2) in our samples, exceeding the age cutoff given in \cite{West2009}, though it follows the empirical relation given by \cite{Kiman2021}.

Further, the observed \ha variability in our sample is shown with respect to the derived ages and spectral type in Fig.~\ref{fig-age-variability}, and higher short-term variability is noticed in older late-type stars. We notice that \cite{Kiman2021} has also tried to relate the \ha variability with the age of the stars  (figure-7 of their article), where higher \ha variability is seen for younger objects. However, their sample covers a larger age span (6.5$<$log(age/years)$<$ 9). Moreover, the \ha variability in their sample is not derived from any systematic short-term time series (such as ours) and therefore, it may not be prudent to compare these results.
\begin{figure*}
	\centering
	\includegraphics[angle=0,width=0.9\textwidth]{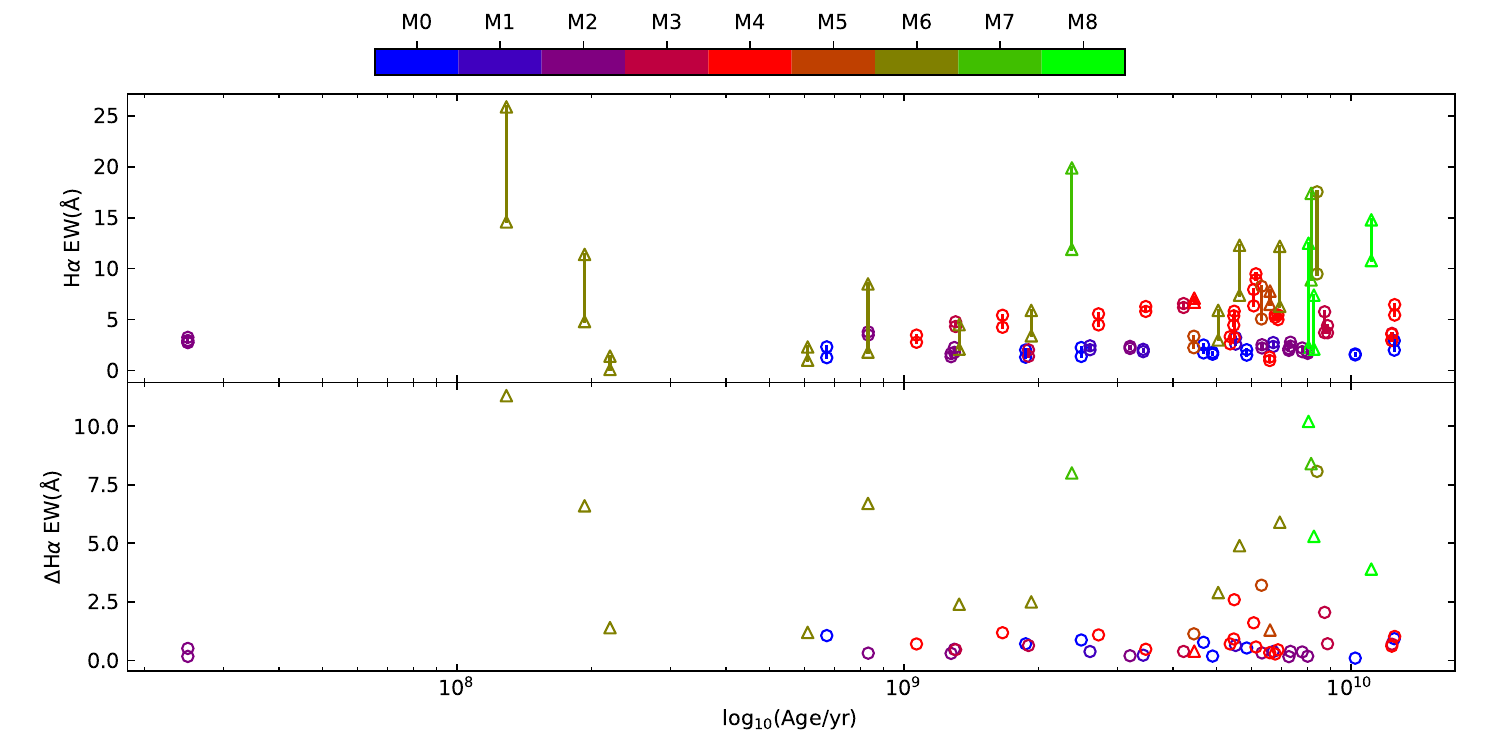}
	\caption{The figure shows \ha variability in our sample (open circle) and \protect\cite{Lee2010} sample (open triangle) as a function of age. The top panel shows the \ha EWs for the sources. The maximum and minimum of the \ha EWs in the time series for a star are connected with a line. The bottom panel shows $\Delta$ EW in the time series.}
	\label{fig-age-variability}
\end{figure*}


\section{Discussion}
\label{Sec-Discussion}

In this work, we have studied the short-term activity of M dwarfs through \ha variability as a proxy. Though our sources belong to M0-M6.5 spectral classes, a similar study by \cite{Lee2010} provided a complementary set of 43 M dwarfs in mostly M3.5-M8.5 spectral types. Thus this set of 126 sources spanning a full range of M0-M8.5 is suitable for exploring the distribution of derived variability across the spectral types. The availability of photometric data from the TESS and Kepler/K2 missions also proved extremely useful for relating the derived short-term spectroscopy variability with their rotation periods and age.\\

We present the results of our analysis as follows:
\par
\begin{enumerate}

\item In our data set of 126 sources, the activity strengths are determined to be close to $\sim10^{-3.8}$ for the spectral types M0-M4 and then decrease to $\sim10^{-5.0}$ for mid to late-type M dwarfs. The derived mean activity strengths ($\langle$L$_{H\alpha}$/L$_{bol}$$\rangle$ and $\langle$L$_{H\beta}$/L$_{bol}$$\rangle$) over this short-term time series are consistent with the trend seen in \cite{Bell2012, Lee2010}, where the activity strengths decrease for the later spectral types, and the corresponding variability was found to be higher.  
	
\item We studied the variability in \ha and \hb using various variability indicators. A noticeable difference in these indicators is seen between early-type (M0-M3) and late-type ($>$ M3) M dwarfs, which is partially due to the intrinsic variability of \ha \citep{Kruse2010, Bell2012}. These breaks in the activity strength at spectral type M3, could be explained due to a change in the magnetic dynamo mechanism at the fully convective boundary.
	
\item Using the cumulative normalized-structure function, we investigated the timescales of variability in \ha EW time series. The observed \ha variability time scale for early M dwarfs (M0-M2) is found to be $\sim$40 - 60 minutes, whereas, for mid-type M dwarfs (M3-M5), it is found to be longer than 60 minutes. This trend is also seen by \citep{Kruse2010}, but we noticed that their observed spectra are not in good cadence. We speculate that different \ha variability time scales could be related to the varying strength of magnetic activity as a function of spectral type.
	
\item Using the high cadence photometric data from TESS and Kepler/K2 missions, we derived the rotation period of 82 M dwarfs in our sample, which ranges between $\sim$0.2-10 days. We find that late-type M dwarfs (M6-M8.5) with a rotation period of $<$ $\sim$2 days show higher variability. This behaviour could be caused by the faster-rotating stars as they are seen to demonstrate magnetic activity in shorter time scales, and it is believed that active dynamo is responsible for this phenomenon \citep{Wright2011}. While the observed variabilities in H-alpha can be the base variabilities due to the appearance and disappearance of starspots or pledges on them \citep{Medina2022}, it is their fast-rotating nature that leads us to believe that we are capturing part of their active phases when they are flaring more and showing the variability in H-alpha. It is also to be mentioned that surface manifestation of higher magnetic activity (such as starspots) will lead to more flares. However, to establish our conjecture, we will require long-time systematic observations of the same objects.
	
\item To determine the rotation-activity relation, we have calculated the Rossby number (R$_0$). We find that (L$_{H\alpha}$/L$_{bol}$ maintains the saturated value till R$_0$ $<$0.2. With rotation above a certain threshold, as suggested by \cite{Newton2017} and \cite{Wright2018}, magnetic activity maintains a constant value, while at slower spins, magnetic activity and rotation are correlated.
	
\item Though the total duration of our spectral time series is only a small fraction of the rotation period (0.4-24\%), for most of the sources, we attempted to check if any correlation could be observed between \ha EW and the rotation phase. We notice apparent correlation/anti-correlation in some of the sources, which in some studies \citep{Baliunas1988, Buccino2011, Fuhrmeister2019, Medina2022} is interpreted as the presence of spots/bright faculae. However, considering the smaller covering fraction of the phase, we restrict ourselves from interpreting this result.
	
\item Using kinematics and parallaxes from GAIA DR3, we could estimate the age of 63 M dwarfs. Though the ages of M dwarfs in our sample range from 0.025-12.528 Gyr, from our analysis, we find that older late-type M dwarfs show higher variability in \ha.

\item The sources having smaller filling factors were found to be more variable across the spectral range. However, the late-type sources are more prominent in the variability scale. A large filling factor could give rise to a strong and persistent \ha emission as opposed to the objects having low filling factors. Any minor change (such as the appearance of new small active regions, micro-flare, etc.) on the highly active stars with high filling factors would not be as prominent as for the stars with low filling factors (lower \ha emission). Thus, stars with smaller filling factors are expected to be more variable, as we notice in our analysis. The same inference has also been drawn by \cite{Bell2012}.

\end{enumerate}

\par 
Variability in M dwarfs on the shorter time scales (a few minutes) is very significant as they are postulated to originate from the small energetic events or the appearance/disappearances of starspots on the stellar surfaces \citep{Bell2012, Medina2022}. Such physical origins of the observed variability in H-alpha are inherently related to the dynamics of stellar magnetic fields, which are known to be coupled with stellar rotation. Thus, studies of variabilities at these short time scales could probe the possible link between activity and rotation. Such studies are beneficial for detecting the low-amplitude and short-duration flares, which can further explore our understanding of flare behavior and their relation with rotational phases (if any). The time scales of variability in M dwarfs are much dependent on the basic parameters (e.g., rotation, magnetic field strengths, etc.) and dynamics (such as internal dynamos) of the objects. High-activity objects with large filling factors are found to be less variable, possibly due to the requirements of more energetic events (such as large flares) to change their observational status in terms of \ha / \hb strengths. Such energetic events may require larger time scales to build up and/or to evolve. The variability of less active stars, on the other hand, would probably be governed by the low energetic events that could occur on shorter time scales. Such stars would thus be expected to have lesser filling factors. Thus, the measurements of the time scales of the \ha / \hb variations are of profound significance to understanding the underlying activity on the surface of the star. A larger sample with finer time resolution could very well be the key to unlocking the physical mechanisms responsible for such activity.

\section{Acknowledgments}
The research work at the Physical Research Laboratory (PRL) is funded by the Department of Space, Government of India. VK thanks PRL for his Ph.D. research fellowship. We thank Veeresh Singh (PRL) and Rishikesh Sharma (PRL) for the useful discussions. The computations were performed on the HPC resources at PRL. J.G.F-T gratefully acknowledges the grant support provided by Proyecto Fondecyt Iniciaci\'on No. 11220340, and also from ANID Concurso de Fomento a la Vinculaci\'on Internacional para Instituciones de Investigaci\'on Regionales (Modalidad corta duraci\'on) Proyecto No. FOVI210020, and from the Joint Committee ESO-Government of Chile 2021 (ORP 023/2021), and from Becas Santander Movilidad Internacional Profesores 2022, Banco Santander Chile. This research has made use of the SIMBAD database, operated at CDS, Strasbourg, France. This paper includes data collected by the TESS and Kepler mission and acquired from the Mikulski Archive for Space Telescopes (MAST) data archive at the Space Telescope Science Institute (STScI). Funding for the TESS and KEPLER missions is provided by NASA's Science Mission Directorate. STScI is operated by the Association of Universities for Research in Astronomy, Inc., under NASA contract NAS5-26555. This research made use of Lightkurve, a Python package for Kepler and TESS data analysis.
\section{Data Availability}
The MFOSC-P spectroscopic monitoring data may be made available on reasonable request. The corresponding authors may be contacted for that. The TESS and KEPLER data used in this work are available on Mikulski Archive for Space Telescopes (MAST).

\bibliography{ref}{}
\bibliographystyle{mnras}

\includepdf[pages=-]{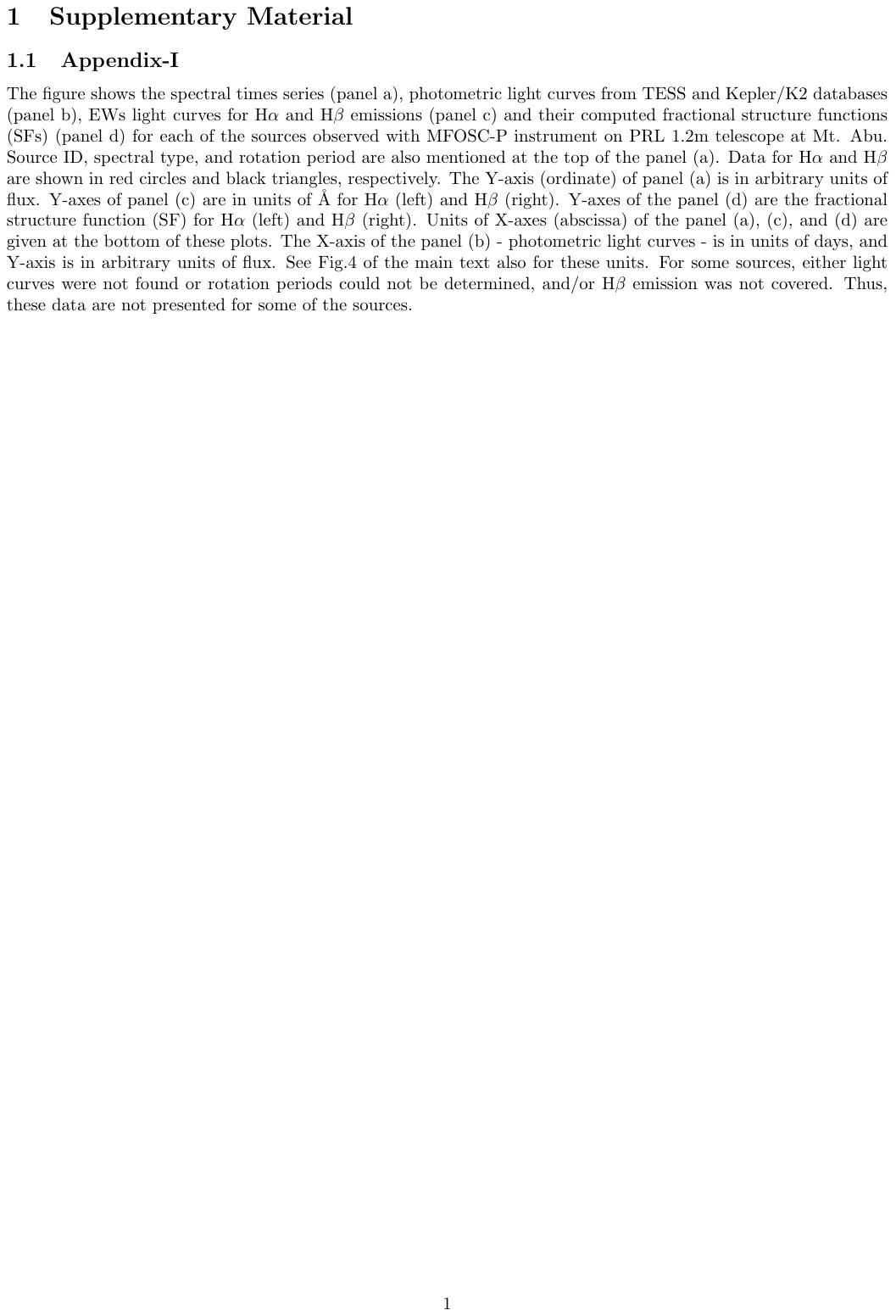}

\end{document}